\documentclass[10pt,newlfont,superscriptaddress,showpacs,twocolumn,aps,pra,letter]{revtex4}
\usepackage{graphicx}
\usepackage[sort&compress]{natbib}
\usepackage{exscale}
\def\hs{\hspace{0.5cm}}

\def\ft{\hspace{0.1cm}}

\def\ea{{\it et al.}}
\def\he4{${}^4$He}

\def\aho{$a_{ho}\,=\,\sqrt{\hbar/(m\omega_{ho})}$}
\def\hw{$\hbar\omega_{ho}$}
\begin{document}
\title{Global and local condensate and superfluid fractions of a 
few hard core Bosons in a combined harmonic optical cubic lattice}
\author{Asaad R. Sakhel}
\affiliation{Al-Balqa Applied University, Faculty of Engineering Technology, 
Applied Sciences Department, Amman 11134, JORDAN}
\begin{abstract}
We explore the global and local condensate and superfluid (SF) 
fractions in a system of a few hard core (HC) bosons ($N=8$ and $N=40$) 
trapped inside a combined harmonic optical cubic lattice (CHOCL) 
at $T=0$ K. The condensate fraction (CF) is computed for individual 
lattice wells by separating the one-body density matrix (OBDM) 
of the whole system into components at the various lattice sites. 
Then each ``lattice-site" component is diagonalized to find its 
eigenvalues. The eigenvalues are obtained by a method presented 
earlier [Dubois and Glyde, Phys. Rev. A {\bf 63}, 023602 (2001)]. 
The effects of interference between the condensates in the lattice 
wells on the CF in one well is also investigated. The SF fraction (SFF)
is calculated for $N=40$ by using the diffusion formula of Pollock 
and Ceperley [Pollock and Ceperley, Phys. Rev. B {\bf 36}, 8343 (1987)]. 
Our chief result is an opposing behavior of the global CF and SFF
with increasing lattice wave vector $k$. In addition, the CF in a 
lattice well is enhanced by the interference with its neighbor wells 
beyond the result when the interference is neglected. The global SF 
is depleted with a rise of the repulsion between the bosons, yet at 
very strong interaction superfluidity is still present. The global
CF remains almost constant with increasing HC repulsion. A reduction 
in the lattice dimension, i.e. an increase in the lattice wave vector, 
increases the local CF in each lattice well, but reduces the corresponding 
local SFF. At large HC repulsion, a coexisting SF-(vacuum)MI phase
is established.
\end{abstract}
\date{\today}
\maketitle

\section{Introduction}

\hs The Bose-Einstein condensation (BEC) of bosons in optical lattices 
(OLs) has recently become a topic of great interest which motivated 
substantial work \cite{Bach:04,Xu:2006,Sun:2009,Chen:2010,Valizadeh:2010,
Ramakumar:2005,Pu:03,Louis:2003,Fabbri:2009,Fang:2010,Yi:2007,
vanOosten:2001}. Interesting investigations included an experimental 
realization of the latter \cite{Bach:04}, BECs in tight binding bands 
of OLs with different geometries \cite{Ramakumar:2005}, the Bose-Hubbard 
model (BHM) \cite{Hen:2010b,Rigol:2009,Hen:2010a,vanOosten:2001}, 
effects of the lattice dimension \cite{Xue:2008}, and instabilities 
of BECs in moving two-dimensional (2D) OLs \cite{Chen:2010}. 

\hs A topic of importance is the measurement of the condensate 
fraction (CF) in individual OL wells, for which only a few 
investigations have been reported \cite{Sun:2009,Diener:07,Brouzos:2010}. 
Investigations concentrated mostly on measuring the {\it total} CF 
of the whole lattice system. For example, in a 2D lattice boson 
system, Spielman \ea\ \cite{Spielman:2008} experimentally measured 
the CF of Rb atoms as a function of the lattice depth. Furthermore, 
Fang \ea\ \cite{Fang:2010} measured the CF of a $^{87}$Rb gas released 
from an OL in a time-of-flight (TOF) experiment.

\hs Importantly, Chen and Wu \cite{Chen:2010} noted that most 
of the theoretical work focused only on BECs in 1D OLs, and that 
the literature on 2D and three-dimensional (3D) OLs is still 
scarce. We were thus motivated to explore BECs in 3D OLs; the 
work of Brouzos \ea\ \cite{Brouzos:2010} provides also the 
chief motivation.

\hs According to Chen and Wu \cite{Chen:2010}, a BEC confined in a 
strong 1D lattice can be regarded as a chain of weakly coupled 
condensate islands trapped in the lattice wells. This idea has also 
been propagated by Shams and Glyde \cite{Shams:09} and there would be 
little tunneling between these islands, as Chen and Wu stated. Our 
upcoming formulation of the present problem is based on the latter 
thought. That is, we study the role of the interference between the 
condensate in one lattice well and the condensates in all-neighbor 
wells. In this regard, we were also motivated by the work of Baillie 
and Blackie \cite{Baillie:2009}.

\hs Further, superfluidity in OLs has also been given 
considerable interest 
\cite{Greiner:02,Gygi:06,Sun:2009,Roth:2003,Diener:07,Tilahun:2011}.
Particularly the superfluid (SF) to Mott-insulator (MI) transition 
\cite{Hen:2009,Sansone:07,Li:06,Hartmann:2008,Roth:2003,Gerbier:05,
Yamashita:07,Capello:07,Ramanan:2009,Rancon:2012} has been explored 
intensively ever since its first experimental realization 
\cite{Greiner:02}, as well as coexisting SF and MI domains of 
harmonically trapped hard core (HC) and lattice bosons 
\cite{Hen:2010a,Rigol:2009,Batrouni:2002,Stoeferle:2004,Ramanan:2009}. 
For example, Roth and Burnet \cite{Roth:2003}, used a twisted-boundary 
condition approach to compute the superfluid fraction (SFF) in a 
one-dimensional (1D) lattice. Similarly, Hen and Rigol \cite{Hen:2009} 
applied a twist in the boundary conditions to evaluate the SFF of HC 
bosons in a superlattice. The twist in the boundary conditions can, 
however, be only applied to a homogeneous OL system. In addition, 
using the BHM, the effects of an external harmonic trap on the state 
diagram of lattice bosons have been explored by Rigol \ea\ 
\cite{Rigol:2009}, specifically for the coexistence of MI and SF 
domains inside the system. The vanishing of the CF at large values 
of the lattice depth indicated also the crossing from the SF to 
the MI domain \cite{Spielman:2008}. 

\hs The role of the HC diameter in the SF and condensate depletion 
of hard sphere (HS) bosons in OLs has, to the best of our knowledge, 
rarely been outlined before, particularly in the SF-MI transition 
of a few-body system. So far, the investigations of the CF and SFF 
have been largely as functions of the lattice barrier (c.f. Shams 
and Glyde \cite{Shams:09} and Spielman \ea\ \cite{Spielman:2008}); 
in contrast we explore them as functions of the boson HC interactions. 
One of the few investigations on the role of the interactions in 
the condensate properties inside an OL was undertaken by Snoek \ea\ 
\cite{Snoek:2011} and Ramanan \ea\ \cite{Ramanan:2009} for a 1D Bose
gas in an inhomogeneous OL. Snoek \ea\ explored the effect of 
interactions on the condensate properties of a Bose-Fermi mixture 
trapped in a 3D harmonic OL. Among their findings, was that the 
condensate is depleted with a rise of the Bose-Fermi repulsive 
interactions which are similar to the boson HC interactions. 
Whereas Snoek \ea\ conducted their calculations for the whole of 
their system, we explore the condensate at individual lattice 
sites and then sum all individual contributions. 

\hs One important investigation most relevant to ours was presented by
Brouzos \ea\ \cite{Brouzos:2010}, who conducted studies on homogeneous
few-particle bosonic systems in a 1D multiwell trap. Some of our 
findings for a 3D OL are similar to theirs. Further, the latter authors 
articulated that exact studies of trapped bosonic systems ``are particular
for a few number of particles". Few particles have also been used, e.g.
by Fang \ea\ \cite{Fang:2010} who did calculations on a 1D Bose gas in
an OL with only ten particles in ten lattice sites. We shall devote a
special section for connecting our results with the findings of 
Ref.\cite{Brouzos:2010} later on below.

\hs In our work here, the CF and SFF of a few HC bosons trapped 
inside a combined harmonic optical cubic lattice (CHOCL) is studied
at $T=0$ K. The cubic OL has $3\times 3\times 3$ lattice sites, or 
lattice wells. Most importantly, the nonspherical symmetry of the 
CHOCL system is dealt with which poses a difficulty for the computation 
of its global CF. This fact has been outlined recently by Astrakharchik 
and Krutitsky \cite{Astrakharchik:2011}, who presented a method for the 
calculation of the total CF of an inhomogeneous Bose gas confined 
by an OL. Yet this work here proposes to evaluate the CF in each 
individual lattice well. Hence, another goal of our paper is to present 
methods for computing the CF and SFF in each well of the CHOCL system. 
The total CF and SFF are also computed. An important point to emphasize, 
is that we distinguish clearly between global and local CF and SFF. 
Motivated by the work of Xue \ea\ \cite{Xue:2008}, we also decided 
to check the role of the lattice dimension on the properties of the 
CF and SFF. In another attempt, we seek a MI state in our systems 
by going to large HC repulsions and OL depths. We also make a clear 
distinction between the effects of repulsive forces with {\it zero 
range}, as encountered in the BHM, and repulsive forces with a 
{\it nonzero} range, as described by the HS Jastrow function, 
Eq.(\ref{eq:Jastrow}) below. A pair of bosons interacting solely by a 
repulsive delta function can still be brought close to each other by 
external forces; but if they are HSs, then their closest distance 
of approach is $a_c$ below which they face an infinite hard-wall barrier.

\hs The current paper comes as a continuation to the investigation 
of the properties of HC bosons in a CHOCL by Sakhel \ea\ 
\cite{Sakhel:2010}, in continuous space and for a few-body system. 
This is in an attempt to provide further manifestation of the 
condensate properties inside an OL. Therefore we ask ourselves 
various questions: Will an increase of the HC diameter deplete 
the SF or BEC inside the CHOCL in the case of a few bosons? Can 
we realize a MI by increasing the HC diameter to large values 
instead of the OL depth? What is the effect of interference between 
the condensates in all lattice wells on the condensate in one well?
We also would like to emphasize that the current investigation 
provides further evidence for the presence of superfluidity in a 
3D CHOCL. Previously, Sun \ea\ \cite{Sun:2009} presented striking 
experimental evidence for the presence of SF states in a shallow OL. 
Sun \ea\ \cite{Sun:2009} found that the CF in a deep lattice is 
significantly lower than 1. Further, in a shallow limit, the SFF 
is 100$\%$ \cite{Sun:2009}. Correspondingly, our findings here 
indicate an overall CF of the order of $\sim 100\%$ in the 
weakly-interacting regime for $N=8$ particles and an OL of depth  
$V_0=10$ (in trap units). On the other hand, the SFF is $\sim 90\%$ 
in the weakly-interacting regime for $V_0=10$ and $N=40$ particles 
(see Fig.\ft\ref{fig:plot.vpi.rho_sf.vs.ac.diffusion.N40.B10.severalk}).

\hs The key findings of this paper are as follows: i) the most 
important result is an opposing behavior of the global CF and 
SFF as functions of the lattice spacing. Whereas a reduction 
in the lattice spacing decreases the global SFF, it leads at 
the same time to an increase in the CF of the whole system. 
This is counter-intuitive and arises from a distinction between 
local condensate density and global SF density; ii) the SFF is 
reduced with a rise of the HC diameter in correspondence to a 
decline in the single-particle tunneling, earlier reported by 
Sakhel \ea; iii) the interference between the condensates in 
all lattice wells enhances the CF in each well beyond the 
result without interference effects; iv) the energy rises with 
increasing HC diameter whereas the SFF declines. As a result, 
one can conclude that the rise in the energy is mostly due to 
a buildup of the onsite repulsive energy in each lattice well 
which overtakes the drop in the boson mobility; v) for a small 
number of HC bosons $N$ inside a CHOCL with a limited number of 
lattice sites $N_L$, it is possible to achieve a mixed SF-MI 
state. vi) the local and global CF, and SFF, are distinct 
quantities. vii) the principle factor in depleting a BEC in a 
CHOCL with few bosons is the OL.

\hs The paper is organized as follows. In Sec.\ref{sec:method}, 
we outline the methods used for the evaluation of the CF and SFF. 
In Sec.\ref{sec:resanddis}, we present our results and discuss 
them, and in Sec.\ref{sec:conclusions} we present our conclusions.

\section{Method}\label{sec:method}

\hs In this section, we describe the methods used in the 
evaluation of the local and global CF and SFF in each lattice 
well, as well as the global CF and SFF. We consider $N$ HC bosons 
confined by a CHOCL of $3\times 3\times 3$ sites. The local CF 
is computed by separating the one-body density matrix (OBDM) 
of the whole system into its components centered at the 
various lattice sites. By treating the distribution of 
particles in each lattice well as a spherically symmetric 
cloud (we show justification for this), we diagonalize the 
OBDM components and find their corresponding eigenvalues for 
a particular number of sites representative of the whole OL. 
This is performed for two cases, one involving no interference 
between the condensate clouds in all lattice wells, and the other 
with this interference included. The global CF is then obtained
by summing the contributions from individual lattice wells. The 
local SFF is evaluated by dividing the large cubic OL volume 
into 27 small cubes of edge $d$, and computing the SFF within 
the boundaries of each small cube. The global SFF is computed 
as well, and for both local and global SFF a winding-number like 
formula is applied.

\hs For the purpose of evaluating the OBDM and the associated 
eigenvalues, we modified a previously written code by Dubois 
and Glyde \cite{DuBois:01}. For the evaluation of the SFF, we 
applied the diffusion formula of Pollock and Ceperley \cite{Pollock:1987}. 
The variational path integral Monte Carlo (VPI) method 
\cite{Dubois:2007,Cuervo:05} was used to evaluate the SFF of 
the systems treated in Ref.\cite{Sakhel:2010}. We used quantum 
variational Monte Carlo (VMC) to compute the spatial 
configurations of the particles in the current systems from 
which the OBDMs are obtained.

\subsection{Hamiltonian}

\hs The Hamiltonian of the system is given by

\begin{eqnarray}
&&H\,=\,-\frac{\hbar^2}{2 m}\sum_{i=1}^N \nabla_i^2\,+\,
\sum_{i=1}^N \left[V_{ho}(\mathbf{r}_i)\,+\,V_{opt}(\mathbf{r}_i)\right]\,+\,
\nonumber\\
&&\sum_{i < j} V_{int}(\mathbf{r}_i\,-\,\mathbf{r}_j),
\end{eqnarray}

where $V_{ho}(\mathbf{r}_i)\,=\,\frac{1}{2} m\omega_{ho}^2 r_i^2$ is 
the harmonic oscillator (HO) trapping potential, with 
$\mathbf{r}_i \equiv (x_i, y_i, z_i)$ the position of a particle 
from the center of this trap, $m$ the mass of the particle, and 
$\omega_{ho}$ the trapping frequency. Upon this trap, there is 
superimposed an OL potential

\begin{equation}
V_{opt}(\mathbf{r}_i)\,=\,V_0\,\left[\sin^2(k_x x_i)\,+\,\sin^2(k_y y_i)\,+
\,\sin^2(k_z z_i)\right],
\end{equation}

with $V_0$ the height of the OL barrier, and $k_i\,=\,\pi/d$ 
($i\equiv x,y,$ or $z$) is the lattice wave vector with $d$ the lattice 
spacing. For further details refer to Ref.\cite{Sakhel:2010}. The 
interparticle interactions are given by the hard sphere (HS) potential

\begin{equation}
V_{int}(r_{ij})\,=\,\left\{\begin{array}{r@{\quad ;\quad}l}
\infty & r_{ij} \le a_c\\
0 & r_{ij} > a_c
\end{array}\right.,
\end{equation}

where $a_c$ is the HC diameter of the bosons, and 
$r_{ij} \equiv |\mathbf{r}_i\,-\,\mathbf{r}_j|$ is the interparticle
distance between two bosons $i$ and $j$. $a_c$ equals the s-wave 
scattering length in the low-energy and long-wavelength 
approximation of the two-particle scattering problem.

\subsection{Density matrix and CF at a lattice site}
\label{sec:density-matrix}

\hs We begin with the VMC trial wave function of a previous 
publication \cite{Sakhel:2010} given by

\begin{eqnarray}
&&\Psi(\{\mathbf{r}\},\{\mathbf{R}\})\,=\,\nonumber\\
&&\prod_{i=1}^N\,\exp(-\alpha r_i^2)\,\psi(\mathbf{r}_i,\{\mathbf{R}\})\,
\prod_{i<j} f(|\mathbf{r}_i-\mathbf{r}_j|),
\label{eq:trial-wave-function}
\end{eqnarray}

where $f(|\mathbf{r}_i-\mathbf{r}_j|)$ is the HS Jastrow function
\cite{DuBois:01}

\begin{equation}
f(r)\,=\,1\,-\,\frac{a_c}{r}, \label{eq:Jastrow}
\end{equation}

with $a_c$ the HC diameter of the bosons. An important 
condition is that $f(r)=0$ if $r \le a_c$. Here 
$\{\mathbf{r}\}\,\equiv\,(\mathbf{r}_1,\mathbf{r}_2,\cdots,\mathbf{r}_N)$
is the set of $N$ particle positions, and 
$\{\mathbf{R}\}\,\equiv\,(\mathbf{R}_1,\mathbf{R}_2,\cdots,\mathbf{R}_{N_L})$
the set of $N_L$ lattice site positions, where 
$\mathbf{R}_n\,\equiv\,(i\mathbf{i}+j\mathbf{j}+k\mathbf{k})\pi/d$ 
and is abbreviated $\mathbf{R}_n\,\equiv\,(ijk)$. The index $n$ 
runs from 1 to $N_L$, where $N_L$ is the total number of lattice 
sites (here 27). Eq.(\ref{eq:trial-wave-function}) is then optimized 
with respect to its parameters [$\alpha$ of Eq.(\ref{eq:trial-wave-function}) 
above, and $\beta$, $\gamma$, and $\sigma$ of Eq.(\ref{eq:phi_similar_to_Li}) 
below] as outlined previously \cite{Sakhel:2010}. By using the 
Wannier-like function defined by

\begin{equation}
\psi(\mathbf{r}_i,\{\mathbf{R}\})\,=\,\sum_{n=0}^{N_L} 
\phi(\mathbf{r}_i,\mathbf{R}_n),
\label{eq:Wannier-like-function}
\end{equation}

with $\phi(\mathbf{r}_i,\mathbf{R}_n)$ given by

\begin{eqnarray}
\phi(\mathbf{r}_i,\mathbf{R}_n)\,&=&\,\exp[-\beta(\mathbf{r}_i-\mathbf{R}_n)^2]
\times\,\nonumber\\
&&\left[1+\gamma(x_i-X_n)^2-\sigma(x_i-X_n)^4\right]\times \nonumber\\
&&\left[1+\gamma(y_i-Y_n)^2-\sigma(y_i-Y_n)^4\right]\times \nonumber\\
&&\left[1+\gamma(z_i-Z_n)^2-\sigma(z_i-Z_n)^4\right], 
\label{eq:phi_similar_to_Li}
\end{eqnarray}

we consider expanding the OBDM into its components at each lattice 
site of position $\mathbf{R}_n$. 

\hs To set the stage, we start out by evaluating the total density 
matrix 

\begin{eqnarray}
&&\rho(\mathbf{r}_1,\mathbf{r}_1^\prime)\,=\,
\int d\mathbf{r}_2 d\mathbf{r}_3 \cdots d\mathbf{r}_N 
\Psi^\star (\mathbf{r}_1,\mathbf{r}_2,\cdots,\mathbf{r}_N,\{\mathbf{R}\}) \times 
\nonumber\\
&&\Psi (\mathbf{r}_1^\prime,\mathbf{r}_2,\cdots,\mathbf{r}_N,\{\mathbf{R}\}),
\end{eqnarray}

using the standard Monte Carlo integration approach \cite{Kalos:86} 

\begin{eqnarray}
\rho(\mathbf{r}_1,\mathbf{r}_1^\prime)\,&=&\,
\frac{1}{P}\sum_{c=1}^P \Psi^*(\mathbf{r}_1,\mathbf{r}_{c 2},
\cdots,\mathbf{r}_{c N},\{\mathbf{R}\})\times\nonumber\\
&&\Psi^*(\mathbf{r}_1^\prime,\mathbf{r}_{c 2},\cdots,
\mathbf{r}_{c N},\{\mathbf{R}\})\frac{1}{w_c},
\label{eq:total-density-matrix}
\end{eqnarray}

where $w_c$ is a configurational weight to be determined later, and
$P$ is the number of Monte Carlo configurations. The subscript $c$ in
$\mathbf{r}_{ci}$ labels the configuration to which particle $i$
belongs. Since according to Eq.(\ref{eq:Wannier-like-function}) 
$\Psi(\{\mathbf{r}\},\{\mathbf{R}\})$ is a sum over all lattice 
sites, we can expand $\rho(\mathbf{r}_1,\mathbf{r}_1^\prime)$ 
[Eq.(\ref{eq:total-density-matrix})] into a sum of components at
positions $\mathbf{R}_n$. The latter sum $\rho(\mathbf{r}_1,\mathbf{r}_1^\prime)$ 
involves interference components between one at $\mathbf{R}_q$ and 
all other lattice sites $\mathbf{R}_{n\ne q}$. In this study we are 
concerned with both interfering and noninterfering ones. On substituting 
(\ref{eq:trial-wave-function}) and (\ref{eq:Wannier-like-function}) 
into (\ref{eq:total-density-matrix}), one gets

\begin{eqnarray}
&&\rho(\mathbf{r}_1^\prime,\mathbf{r}_1)\,=\,
e^{-\alpha r_1^2} e^{-\alpha r_1^{\prime 2}} \times\nonumber\\
&&\left[\sum_{n=1}^{N_L} \phi(\mathbf{r}_1,\mathbf{R}_n)\right]
\left[\sum_{n=1}^{N_L} 
\phi(\mathbf{r}_1^\prime,\mathbf{R}_n)\right]\times
\nonumber\\
&&\frac{1}{P}\sum_{c=1}^P \frac{1}{w_c}\prod_{1<j} 
f(|\mathbf{r}_1-\mathbf{r}_{c j}|)
\prod_{1<j} f(|\mathbf{r}_1^\prime-\mathbf{r}_{c j}|) \times \nonumber\\
&&\left\{\prod_{i\ne 1}^N e^{-\alpha r_{c i}^2} \left[ \sum_{n=1}^{N_L} 
\phi(\mathbf{r}_{c i},\mathbf{R}_n)\right] \prod_{i<j, i\ne1}^N 
f(|\mathbf{r}_{c i}-\mathbf{r}_{c j}|)\right\}^2. \nonumber\\
\label{eq:total-density-matrix-in-detail}
\end{eqnarray}

Let us now choose the configurational weight to be of the form

\begin{eqnarray}
&&w_c\,=\,|\Psi[(\mathbf{r}_1,\mathbf{r}_{c2},\cdots,\mathbf{r}_{cN}),\{\mathbf{R}\}]|^2\,
\nonumber\\
&&=e^{-2\alpha r_1^2}\left[\sum_{n=1}^{N_L} 
\phi(\mathbf{r}_1,\mathbf{R}_n)\right]^2
\left[\prod_{1<j} f(|\mathbf{r}_1-\mathbf{r}_{c j}|)\right]^2 \times
\nonumber\\
&&\prod_{i\ne1}^N e^{-2\alpha r_{c i}^2}\left[\sum_{n=1}^{N_L} 
\phi(\mathbf{r}_{c i},\mathbf{R}_n) \right]^2
\left[\prod_{i<j,i\ne 1} f(|\mathbf{r}_{c i}-\mathbf{r}_{c j}|)\right]^2,
\nonumber\\
\label{eq:total-configurational-weight}
\end{eqnarray}

which has been chosen in a previous publication \cite{Sakhel:2010}.
Substituting this weight into (\ref{eq:total-density-matrix-in-detail}),
the exponentials and Jastrow terms $f(|\mathbf{r}_{c i}-\mathbf{r}_{c j}|)$
cancel out, and one remains with

\begin{eqnarray}
&&\rho(\mathbf{r}_1^\prime,\mathbf{r}_1)\,=\,
\frac{e^{-\alpha r_1^{\prime 2}}}{e^{-\alpha r_1^2}} \times\nonumber\\
&&\frac{\left[\sum_{n=1}^{N_L} 
\phi(\mathbf{r}_1,\mathbf{R}_n)\right]\left[\sum_{n=1}^{N_L} 
\phi(\mathbf{r}_1^\prime,\mathbf{R}_n)\right]}{\left[\sum_{n=1}^{N_L} 
\phi(\mathbf{r}_1,\mathbf{R}_n)\right]^2}
\times \nonumber\\
&&\frac{1}{P}\,\sum_{c=1}^P\,
\frac{\displaystyle\prod_{1<j} f(|\mathbf{r}_1^\prime-\mathbf{r}_{c j}|)}
{\displaystyle\prod_{1<j} f(|\mathbf{r}_1-\mathbf{r}_{c j}|)}.
\label{eq:total-density-matrix-weighted}
\end{eqnarray}

\subsubsection{Interfering and noninterfering components of the OBDM}

\hs In what follows, we shall separate the density matrix 
$\rho(\mathbf{r}_1,\mathbf{r}_1^\prime)$ into interfering and 
noninterfering (isolated-islands) components. First, the numerator
in the second line of Eq.(\ref{eq:total-density-matrix-weighted}) 
is rewritten

\begin{eqnarray}
&&\left[\sum_{n=1}^{N_L} \phi(\mathbf{r}_1,\mathbf{R}_n)\right]
\left[\sum_{q=1}^{N_L} 
\phi(\mathbf{r}_1^\prime,\mathbf{R}_q)\right]\,=\,\nonumber\\
&&\sum_{n=1}^{N_L}\left[\phi(\mathbf{r}_1,\mathbf{R}_n)
\phi(\mathbf{r}_1^\prime,\mathbf{R}_n)\,+\,
\sum_{q\ne n}^{N_L}\phi(\mathbf{r}_1,\mathbf{R}_n)
\phi(\mathbf{r}_1^\prime,\mathbf{R}_q)\right],\nonumber\\
\label{eq:numerator-sum}
\end{eqnarray}

where the first term in the second line of (\ref{eq:numerator-sum})
is the noninterfering ($n=q$) part describing isolated clouds, 
and the second term the interfering part ($q\ne n$), respectively. 
Next, by using (\ref{eq:numerator-sum}) in 
(\ref{eq:total-density-matrix-weighted}), 
the total density matrix can be written as the sum of two components

\begin{eqnarray}
&&\rho(\mathbf{r}_1,\mathbf{r}_1^\prime)\,=\,
\sum_{n=1}^{N_L} 
\rho^{(0)}(\mathbf{r}_1\,-\,\mathbf{R}_n,\mathbf{r}_1^\prime\,-\,\mathbf{R}_n)\,+\,
\nonumber\\
&&\sum_{n=1}^{N_L}\rho^{(1)}
(\mathbf{r}_1\,-\,\mathbf{R}_n,\mathbf{r}_1^\prime\,-\,\mathbf{R}_n),
\label{eq:density-matrix-components}
\end{eqnarray}

where we define the noninterfering density matrix for each ``isolated" 
cloud at a lattice site $\mathbf{R}_n$ as

\begin{eqnarray}
&&\rho^{(0)}(\mathbf{r}_1-\mathbf{R}_n,
\mathbf{r}_1^\prime-\mathbf{R}_n)\,=\,\nonumber\\
&&e^{-\alpha r_1^{\prime 2}} e^{\alpha r_1^2} 
\frac{\displaystyle\phi(\mathbf{r}_1,\mathbf{R}_n) 
\phi(\mathbf{r}_1^\prime,\mathbf{R}_n)}
{\displaystyle\left[\sum_{q=1}^{N_L} \phi(\mathbf{r}_1,\mathbf{R}_q)\right]^2}
J(\mathbf{r}_1^\prime,\mathbf{r}_1),
\label{eq:density-matrix-for-each-lattice-site}
\end{eqnarray}

and the density matrix involving interference components
only ($n\ne q$) as 

\begin{eqnarray}
&&\rho^{(1)}(\mathbf{r}_1\,-\,\mathbf{R}_n,
\mathbf{r}_1^\prime\,-\,\mathbf{R}_n)\,=\,\nonumber\\
&&e^{-\alpha r_1^{\prime 2}}\,e^{\alpha r_1^2}\,
\frac{\phi(\mathbf{r}_1,\mathbf{R}_n)\,
\sum_{q\ne n}^{N_L} \phi(\mathbf{r}_1^\prime,\mathbf{R}_q)}
{\displaystyle \left[\sum_{q=1}^{N_L}\phi(\mathbf{r}_1,\mathbf{R}_q)\right]^2}
J(\mathbf{r}_1^\prime,\mathbf{r}_1).\nonumber\\
\label{eq:interference-density-matrix}
\end{eqnarray}

For brevity, we have defined the term involving Jastrow functions
as

\begin{equation}
J(\mathbf{r}_1^\prime,\mathbf{r}_1)\,=\,
\frac{1}{P}\sum_{c=1}^P \frac{\prod_{1<j} 
f(|\mathbf{r}_1^\prime\,-\,\mathbf{r}_{c j}|)}
{\prod_{1<j} f(|\mathbf{r}_1-\mathbf{r}_{c j}|)}.
\end{equation}

Note that the order in which the sums of $\phi(\mathbf{r}_i,\mathbf{R}_n)$ 
in Eq.(\ref{eq:total-density-matrix-weighted}) are multiplied 
and divided is very important to obtain 
Eqs.(\ref{eq:density-matrix-for-each-lattice-site})
and (\ref{eq:interference-density-matrix}). First, we evaluated 
the numerator (\ref{eq:numerator-sum}) by extracting out the 
interfering $(n\ne q)$ and noninterfering terms ($n=q$). Second,
we divided by the denominator 
$\left[\sum_{n=1}^{N_L}\phi(\mathbf{r}_1,\mathbf{R}_n)\right]^2$. Yet
if we considered first canceling out the summation 
$\sum_{n=1}^{N_L}\phi(\mathbf{r}_1,\mathbf{R}_n)$ in the
numerator of the second line of Eq.(\ref{eq:total-density-matrix-weighted}), 
with one power of the sum in the denominator, we would get a different result 
given by Eq.(\ref{eq:density-matrix-interference}) next. 

\subsubsection{Eigenvalues of the CHOCL-OBDM components}

\hs Assuming now the cloud at each lattice site to be spherical, 
we utilize in what follows the recipe of Dubois and Glyde 
\cite{DuBois:01}, which is for the evaluation of the CF in an 
isotropic harmonic trap, to compute the CF in each lattice well. 
Let us first consider the noninterfering case. First, one begins 
by expanding the density matrix 
(\ref{eq:density-matrix-for-each-lattice-site}) at lattice site 
$\mathbf{R}_n$ into angular momentum components as follows:

\begin{eqnarray}
&&\rho^{(0)}(\mathbf{r}_1-\mathbf{R}_n,\mathbf{r}_1^\prime-\mathbf{R}_n)\,=\,
\nonumber\\
&&\sum_{\ell=0}^\infty \frac{(2\ell+1)}{4\pi} 
P_\ell\left(u_{\mathbf{r}_1,\mathbf{r}_1^\prime}^{(n)}\right)
\rho_\ell^{(0)}(|\mathbf{r}_1-\mathbf{R}_n|,
|\mathbf{r}_1^\prime-\mathbf{R}_n|),\nonumber\\
\label{eq:rho0-noninterfering}
\end{eqnarray}

where $P_\ell(x)$ is the Legendre polynomial of order $\ell$ and

\begin{equation}
u_{\mathbf{r}_1,\mathbf{r}_1^\prime}^{(n)}\,=\,
\frac{(\mathbf{r}_1-\mathbf{R}_n)}{|\mathbf{r}_1-\mathbf{R}_n|}
\cdot\frac{(\mathbf{r}_1^\prime-\mathbf{R}_n)}
{|\mathbf{r}_1^\prime-\mathbf{R}_n|},
\end{equation}

the cosine of the angle between ($\mathbf{r}_1-\mathbf{R}_n$)
and ($\mathbf{r}_1^\prime-\mathbf{R}_n$). The $\rho_\ell^{(0)}$
corresponds to the angular momentum component of $\rho^{(0)}$
and is obtained by multiplying both sides of 
(\ref{eq:rho0-noninterfering}) by 
$P_\ell(u^{(n)}_{\mathbf{r}_1,\mathbf{r}_1^\prime})$ and 
integrating over the solid-angle element 
$d\Omega_1\,=\,2\pi \sin\gamma d\gamma$, considering that
$\cos\gamma\,=\,u^{(n)}_{\mathbf{r}_1,\mathbf{r}_1^\prime}$.
This yields

\begin{eqnarray}
&&\rho_\ell^{(0)}(|\mathbf{r}_1-\mathbf{R}_n|,
|\mathbf{r}_1^\prime-\mathbf{R}_n|)\,=\,\nonumber\\
&&\int d\Omega_1 P_\ell\left(u_{\mathbf{r}_1,\mathbf{r}_1^\prime}^{(n)}\right)
\rho^{(0)}(\mathbf{r}_1-\mathbf{R}_n,\mathbf{r}_1^\prime-\mathbf{R}_n),
\nonumber\\
\label{eq:density-matrix-for-each-angular-momentum}
\end{eqnarray}

where the orthogonality condition for Legendre polynomials 
\cite{Arfken:1995} has been applied:

\begin{equation}
\int_0^\pi P_\ell(\cos\gamma) P_{\ell^\prime}(\cos\gamma)
\sin\gamma d\gamma\,=\,\frac{2\delta_{\ell\ell^\prime}}{2\ell\,+\,1},
\label{eq:Legendre-polynomial-orthogonality-condition}
\end{equation}

with $\delta_{\ell\ell^\prime}$ the Kronecker delta function.

\hs Second, one defines the local natural orbitals (LNOs) and 
diagonalizes the local OBDM to evaluate the eigenvalues of the
OBDM in each well. Hence, similarly to Dubois and Glyde 
\cite{DuBois:01} one first redefines the local noninterfering 
density matrix (\ref{eq:rho0-noninterfering}) using their 
field-operator approach:

\begin{equation}
\rho^{(0)}(\mathbf{r}_1\,-\,\mathbf{R}_n,\mathbf{r}_1^\prime-\mathbf{R}_n)\,=\,
\langle\hat{\Psi}_0^\dagger(\mathbf{r}_1-\mathbf{R}_n) 
\hat{\Psi}_0(\mathbf{r}_1^\prime-\mathbf{R}_n)\rangle,
\label{eq:local-obdm-field-operator-approach}
\end{equation}

$\hat{\Psi}_0(\mathbf{r}_1-\mathbf{R}_n)$ being a local field 
operator which annihilates a particle at position 
$\mathbf{r}_1-\mathbf{R}_n$. Then one expands 
$\hat{\Psi}_0(\mathbf{r}_1-\mathbf{R}_n)$ into a set of local 
single particle states $\chi_i(\mathbf{r}_1-\mathbf{R}_n)$ around 
$\mathbf{R}_n$:

\begin{equation}
\hat{\Psi}_0(\mathbf{r}_1-\mathbf{R}_n)\,=\,
\sum_{i=1}^N \chi_i(\mathbf{r}_1-\mathbf{R}_n)\,\hat{a}_i^{(n)},
\label{eq:psi-expanded-into-single-particle-states}
\end{equation}

$\hat{a}_i^{(n)}$ being a bosonic annihilation operator acting
inside well $n$ at position $\mathbf{R}_n$. The $\chi_i$ are 
taken to be orthonormal such that

\begin{equation}
\int d\mathbf{r}_1 \chi_i^*(\mathbf{r}_1-\mathbf{R}_n)
\chi_j(\mathbf{r}_1-\mathbf{R}_n) \,=\,\delta_{ij}.
\end{equation}

Subsituting
(\ref{eq:psi-expanded-into-single-particle-states}) into 
(\ref{eq:local-obdm-field-operator-approach}) above, and
using the usual condition
$\langle \hat{a}_i^{(n) \dagger} \hat{a}_j^{(n)}\rangle\,=\,N_i^{(n)}\delta_{ij}$,
one gets

\begin{equation}
\rho^{(0)}(\mathbf{r}_1-\mathbf{R}_n,\mathbf{r}_1^\prime-\mathbf{R}_n)\,=\,
\sum_{ij} \chi_i^*(\mathbf{r}_1-\mathbf{R}_n) 
\chi_j(\mathbf{r}_1^\prime-\mathbf{R}_n) N_i^{(n)} \delta_{ij},
\label{eq:rho0-expanded-into-single-particle-states}
\end{equation}

Multiplying both sides of 
(\ref{eq:rho0-expanded-into-single-particle-states}) by 
$\chi_j(\mathbf{r}_1-\mathbf{R}_n)$ from the left and
$\chi_i^*(\mathbf{r}_1^\prime-\mathbf{R}_n)$ from the right,
then integrating over $\mathbf{r}_1$ and $\mathbf{r}_1^\prime$,
this yields

\begin{eqnarray}
&&\int d\mathbf{r}_1 d\mathbf{r}_1^\prime
\chi_j(\mathbf{r}_1-\mathbf{R}_n) 
\rho^{(0)}(\mathbf{r}_1-\mathbf{R}_n,\mathbf{r}_1^\prime-\mathbf{R}_n)
\times\nonumber\\
&&\chi_i^*(\mathbf{r}_1^\prime-\mathbf{R}_n)\,=\,N_i \delta_{ij}.
\end{eqnarray}

The eigenvectors $\chi_i(\mathbf{r}_1-\mathbf{R}_n)$ are the LNOs
with eigenvalues $N_i^{(n)}$. Making now use of the spherical 
symmetry of the system, one further defines

\begin{equation}
\chi_i(\mathbf{r}_1-\mathbf{R}_n)\,=\,
\xi_{q\ell}(|\mathbf{r}_1-\mathbf{R}_n|) Y_{\ell m}(\theta,\varphi),
\label{eq:chi-radial-spherical-harmonic}
\end{equation}

$i\equiv(q\ell m)$ being a state, 
$\xi_{q\ell}(|\mathbf{r}_1-\mathbf{R}_n|)$ being the local 
radial wave function, and $Y_{\ell m}(\theta,\varphi)$ the
spherical harmonic function with $\theta$ and $\varphi$ 
defining the angular position of $\mathbf{r}_1-\mathbf{R}_n$. 
It must be noted, that in $N_i^{(n)}$ we drop the quantum
number $m$ so that $N_i^{(n)}\,=\,N_{q\ell}^{(n)}$ because of
azimuthal symmetry in each well. Substituting 
(\ref{eq:chi-radial-spherical-harmonic}) and 
(\ref{eq:rho0-noninterfering}) into 
(\ref{eq:rho0-expanded-into-single-particle-states}), using the
addition theorem for Legendre polynomials \cite{Arfken:1995}

\begin{equation}
P_\ell(\cos\gamma)\,=\,\frac{4\pi}{2\ell+1}\sum_{m=-\ell}^{m=+\ell}
Y_{\ell m}(\theta,\varphi) Y_{\ell m}^*(\theta^\prime,\varphi^\prime)
\label{eq:addition-theorem-for-Legendre-polynomials}
\end{equation}

with 

\begin{displaymath}
\cos\gamma\,=\,
\sin\theta \sin\theta^\prime \cos(\varphi-\varphi^\prime)
+\cos\theta \cos\theta^\prime,
\end{displaymath}

and the orthogonality condition 
(\ref{eq:Legendre-polynomial-orthogonality-condition}),
this eventually yields the expansion of the angular 
momentum component (\ref{eq:density-matrix-for-each-angular-momentum}) 
into the radial functions $\xi_{q \ell}$

\begin{eqnarray}
&&\rho_\ell^{(0)}(|\mathbf{r}_1-\mathbf{R}_n|,
|\mathbf{r}_1^\prime-\mathbf{R}_n|)\,=\,\nonumber\\
&&\sum_q \xi_{q \ell}(|\mathbf{r}_1-\mathbf{R}_n|) 
\xi_{q \ell}(|\mathbf{r}_1^\prime-\mathbf{R}_n|)
N_{q \ell}^{(n)},
\label{eq:eigenvalue-natorb}
\end{eqnarray}

where $N_{q \ell}^{(n)}$ are the eigenvalues of the natural orbitals
$\chi_i(\mathbf{r}_1-\mathbf{R}_n)$ at lattice site $n$ and positions 
$\mathbf{R}_n\equiv (ijk)$. As such, the natural orbital 
$\chi_0(\mathbf{r}_1-\mathbf{R}_n)$ whose eigenvalue is $N_{00}^{(n)}$, 
is defined as the condensate orbital. $N_{00}^{(n)}/N$ is then the CF 
at site $n$ and position $\mathbf{R}_n$ with respect to the overall 
number of particles $N$. One then proceeds similarly to Dubois and 
Glyde \cite{DuBois:01} in the diagonalization of the OBDM in order to 
compute the eigenvalues $N_{q \ell}^{(n)}$, and consequently the CF, 
except that we do this here for one lattice site at a time.

\hs By considering the additional interference between a condensate 
at lattice site $\mathbf{R}_n$ and all-neighbor lattice sites 
$\mathbf{R}_q\ne\mathbf{R}_n$, the CF at lattice
site $\mathbf{R}_n$ is computed using the same previous procedure,
except that we use all terms of the density matrix 
(\ref{eq:total-density-matrix-weighted}) which contains the interference 
between the condensates at the various lattice sites, instead of the 
noninterfering $\rho^{(0)}$ only. That is, one writes

\begin{eqnarray}
&&\rho(\mathbf{r}_1-\mathbf{R}_n,\mathbf{r}_1^\prime-\mathbf{R}_n)\,=\,
e^{-\alpha r_1^{\prime 2}}\,e^{\alpha r_1^2}\times\nonumber\\
&&\frac{\phi(\mathbf{r}_1^\prime,\mathbf{R}_n)}
{\left|\sum_{q=1}^{N_L} \phi(\mathbf{r}_1,\mathbf{R}_q)\right|}
J(\mathbf{r}_1^\prime,\mathbf{r}_1), \nonumber\\
\label{eq:density-matrix-interference}
\end{eqnarray}

where $\rho\,=\,\rho^{(0)}\,+\,\rho^{(1)}$. Consequently, 
Eqs.(\ref{eq:rho0-noninterfering}), 
(\ref{eq:density-matrix-for-each-angular-momentum}), and 
(\ref{eq:eigenvalue-natorb}) are simply rewritten by replacing
$\rho^{(0)}$ with $\rho$ in all terms. Eventually, one gets
instead of (\ref{eq:eigenvalue-natorb})

\begin{eqnarray}
&&\rho_\ell(|\mathbf{r}_1-\mathbf{R}_n|,
|\mathbf{r}_1^\prime-\mathbf{R}_n|)\,=\,\nonumber\\
&&\sum_q \widetilde{\xi}_{q \ell}(|\mathbf{r}_1-\mathbf{R}_n|) 
\widetilde{\xi}_{q \ell}(|\mathbf{r}_1^\prime-\mathbf{R}_n|) 
\widetilde{N^{(n)}_{q \ell}},
\end{eqnarray}

where $\widetilde{N^{(n)}_{q \ell}}$ and $\widetilde{\xi}_{q\ell}$ 
are similar to $N^{(n)}_{q \ell}$ and $\xi_{q\ell}$, but including 
contributions from the interference with all-neighbor lattice sites.

\subsection{Definitions of CF in our systems}

\hs To set the stage, we define $n_{(ijk)}\,=\,N^{(n)}_{00}/N$ to 
be the CF in the lattice well centered at $\mathbf{R}_n\equiv(ijk)$ 
with respect to the total number of particles $N$, and {\it without} 
all-well interference, and 
$\tilde{n}_{(ijk)}\,=\,\widetilde{N^{(n)}_{00}}/N$ as the corresponding 
CF {\it with} all-neighbor interference, labelled by a star in each 
upcoming figure legend. The $n_{(ijk)}$ and $\tilde{n}_{(ijk)}$ are 
computed as explained in Sec.\ref{sec:density-matrix}.

\hs Hence $n_{(000)}$ is the fraction of particles in the lowest orbital 
of the central cell, $n_{(010)}$ that of cell $(010)$, etc. In fact, 
the natural orbital is spread out over the entire lattice and is not 
localized in one lattice site. That is, one has a lowest orbital 
at all lattice sites $(ijk)$ if they display a CF. This spreading 
is a Wannier sum over individual-cell natural orbitals. Since the 
CF at site $(ijk)$ is with respect to the total number of particles 
$N$, the total CF of the system in the $3\times 3\times 3$ cubic 
lattice can be computed using

\begin{eqnarray}
&&n_0\,=\,\sum_{ijk} n_{(ijk)}\,\approx\,\nonumber\\
&&n_{(000)}\,+\,6\,n_{(010)}\,+\,12\,n_{(011)}\,+\,8\,n_{(111)}.
\label{eq:total-condensate-fraction}
\end{eqnarray}

Note that $n_{(010)}$ is ideally identical to $n_{(001)}$, $n_{(100)}$,
$n_{(-1 0 0)}$, etc. Similarly $n_{(011)}$ is equal to $n_{(101)}$,
$n_{(110)}$, $n_{(-110)}$, etc; the same holds for $n_{(111)}$. The
large cube divided into $3\times 3\times 3$ smaller equal-sized 
cubes has 6 lattice sites at its face-centers, 8 at the corners, 
and 12 at the bisecting points of its edges.

\hs The CF in each lattice well is then explored as a function 
of $a_c$ for the three lattice wave vectors $k=\pi$, 1.2$\pi$, 
and 1.4$\pi$. The latter results are compared with the corresponding 
CF $\tilde{n}_{(ijk)}$.

\subsection{Superfluid fraction}\label{eq:VPIsuperfluid}

\hs The global SFF is computed using the variational path 
integral Monte Carlo (VPI) technique \cite{Dubois:2007,Cuervo:05} 
via the diffusion formula of Pollock and Ceperely \cite{Pollock:1987}:

\begin{equation}
\frac{\rho_s}{\rho}\,=\,\frac{D_p}{D_0},
\label{eq:super-fluid-fraction-pollock-ceperley}
\end{equation}

where $D_0\,=\,\hbar^2/(2 m)$ is the ``quantum diffusion" 
constant and

\begin{equation}
D_p\,=\,\frac{1}{2 d \beta N}\left\langle
\left[\sum_{i=1}^N(\mathbf{r}_i-
\mathbf{r}_{M i})\right]^2\right\rangle,
\end{equation}

with $d$ the dimensionality of the system, $N$ the number of particles, 
imaginary time $\beta\leftrightarrow 1/(k_B T)$ with $k_B$ the 
Boltzmann constant, and $T$ the temperature. Here $\langle\cdots\rangle$
denotes a configurational Monte Carlo average, $\mathbf{r}_i$ 
is the initial position of a particle $i$, and $\mathbf{r}_{M i}$ 
the destination of the particle after a ``time" $\beta$, where $M$ 
is the number of VPI time slices. In this regard, we evaluate the 
configuational average $D_p$ over a number $P$ of VPI 
configurations:

\begin{equation}
D_{p,\,VPI}\,=\,\frac{1}{2 d \beta N}\frac{1}{P}\sum_{c=1}^P 
\left[\sum_{i=1}^N (\mathbf{r}_{c i}-\mathbf{r}_{M ci})\right]^2.
\end{equation}

For $d=3$, and using units of the trap $a_{ho}=\sqrt{\hbar/(m\omega_{ho})}$
and $\hbar\omega_{ho}$ for length and energy, respectively, the 
SFF is then recast into the form

\begin{equation}
n_s\,=\,\frac{\rho_s}{\rho}\,=\,\frac{1}{3 M\tau}\frac{1}{N P}
\sum_{c=1}^P \left[\sum_{i=1}^N 
(\mathbf{r}_{c i}-\mathbf{r}_{M c i})\right]^2,
\label{eq:VPI-superfluid-fraction}
\end{equation}

where $\mathbf{r}\rightarrow\mathbf{r}/a_{ho}$, 
$\beta\rightarrow\beta\hbar\omega_{ho}$, and 
$\tau\rightarrow\beta\hbar\omega_{ho}/M\,=\,
\hbar\omega_{ho}/(M k_B T)$ is the ``time 
step" with $M$ the number of time slices. We would also like to 
remind the reader that we are aiming at presenting qualitative 
rather than quantitative results. The parameters used were $M=120$, 
$\tau=5\times 10^{-3}$, and $N=40$. The number of MC blocks was 
$\sim O(10^3)$, similarly for the number of MC steps for each block.

\hs In order to compute the local SFFs in the individual CHOCL 
wells using Eq.(\ref{eq:VPI-superfluid-fraction}), we divided 
the large volume of our cubic OL into 27 smaller cubic cells of 
equal sizes $d^3$. The cube edge was equal to the lattice spacing 
$d\,=\pi/k$. The central cell was centered at the origin $(000)$ 
with its faces parallel to the coordinate planes, the rest of the
cells being centered at the other lattice sites. By restricting 
Eq.(\ref{eq:VPI-superfluid-fraction}) to the boundaries of each
cell volume, we were able to compute the SFF $(\rho_s/\rho)_{(ijk)}$ 
in each lattice well. For each cell $(ijk)$, we compute 
$(\rho_s/\rho)_{s (ijk)}$ using only the particles which 
are positioned inside the cell according to the boundary conditions

\begin{equation}
\mathbf{r}\equiv\left\{\begin{array}{l@{\quad : \quad}r}
x & \displaystyle\left(i-\frac{1}{2}\right)\frac{\pi}{k} < x < \left(i+\frac{1}{2}\right)\frac{\pi}{k}, \\
y & \displaystyle\left(j-\frac{1}{2}\right)\frac{\pi}{k} < y < \left(j+\frac{1}{2}\right)\frac{\pi}{k}, \\
z & \displaystyle\left(k-\frac{1}{2}\right)\frac{\pi}{k} < x < \left(k+\frac{1}{2}\right)\frac{\pi}{k}.
\end{array}\right.
\label{eq:SFinsideaCell}
\end{equation}

That is, considering (\ref{eq:SFinsideaCell}) the local SFF is 

\begin{equation}
\left(\frac{\rho_s}{\rho}\right)_{(ijk)}\,=\,\left(\frac{\rho_s}{\rho}\right)
\times \frac{N}{\langle N_{(ijk)}\rangle}.
\label{eq:SFinsideaCell-semi-global-view}
\end{equation}

In this paper, $(\rho_s/\rho)_{(ijk)}$ is computed using the latter
equation, but sometimes it is found more reasonable to compute it 
{\it with respect to the total $N$}, particularly if 
$\langle N\rangle_{(ijk)}\rightarrow 0$ in the local MI phase.
That is, we use (\ref{eq:VPI-superfluid-fraction}) plus 
(\ref{eq:SFinsideaCell}), but without multiplying by the factor 
$N/\langle N_{(ijk)}\rangle$.

\subsection{Optical densities}

\hs In this article, we are also concerned with the average VPI 
integrated optical 2D density, $\langle n_{2D}(x,y)\rangle$. 
The integration is of the total density along the $z-$axis. That is,
we first define a density $n_{2D, c}(x,y)$ for each VPI configuration 
$c$ such that

\begin{equation}
n_{2D, c}(x,y)\,=\,\int_{-\infty}^{+\infty} |\Psi_c(x,y,z)|^2 dz,
\end{equation}

where $\Psi_c(x,y,z)$ is the wavefunction [Eq.(\ref{eq:trial-wave-function})] 
obtained for a certain VPI configuration $c$. Then one takes the 
average over all $P$ configurations

\begin{equation}
\langle n_{2D}(x,y)\rangle\,=\,\frac{1}{P}\sum_{c=1}^P n_{2D, c}(x,y).
\label{eq:n2D}
\end{equation}

Next to this, we also display at some point the 1D optical density,
$\langle n_{1D}(x)\rangle$, which is obtained by integrating the 
2D density along one of the axes:

\begin{equation}
\langle n_{1D}(x)\rangle\,=\,\int_{-\infty}^{+\infty} dy \langle n_{2D}(x,y)\rangle.
\label{eq:n1D}
\end{equation}

We therefore would like to point out, that in Ref.\cite{Sakhel:2010}
we mistakenly wrote that $n_{1D}(x)\,=\,\langle n_{2D}(x,y=0)\rangle$,
whereas it should be as in Eq.(\ref{eq:n1D}). Nevertheless, in the
main text of Ref.\cite{Sakhel:2010} and in its figure captions,
it was clearly indicated that $n_{1D}(x)$ is the integrated
1D optical density, i.e., as in Eq.(\ref{eq:n1D}) above.

\subsection{Units}

\hs As in Ref.\cite{Sakhel:2010} we use units of the trap, \aho\ 
and \hw\ for lengths and energies, respectively, all throughout 
our calculations. That is, $\mathbf{r}\rightarrow \mathbf{r}/a_{ho}$, 
$V_0\rightarrow V_0/(\hbar\omega_{ho})$, $a_c\rightarrow a_c/a_{ho}$, 
$V_{ho}(\mathbf{r})\rightarrow V_{ho}(\mathbf{r})/(\hbar\omega_{ho})
\,=\,(1/2)r^2$, $\mathbf{k}\rightarrow\mathbf{k} a_{ho}$. The optical 
densities $\langle n_{2D}(x,y)\rangle$ and $\langle n_{1D}(x)\rangle$
are in units of $a_{ho}^{-2}$ and $a_{ho}^{-1}$, respectively.

\section{Results and Discussion}\label{sec:resanddis}

\hs In this section, we present the results of our numerical 
calculations. The same systems are considered as in Ref.
\cite{Sakhel:2010}. The latter are $N$ HC bosons confined 
inside a cubic OL of $N_L=3 \times 3 \times 3$ 
lattice sites plus an external harmonic trap. For further 
details about the system and the computational setup, please 
refer to the previous publication \cite{Sakhel:2010}. 

\hs We explore BEC and superfluidity both from a global and a 
local perspective. The local perspective, pertains to their 
fractions in each lattice well (or cell), and the global one 
pertains to these fractions for the whole system (all lattice 
wells). 

\hs From the local perspective, we begin with a brief 
investigation of the ideal Bose gas, in which we show that 
the lattice wells have a low CF for $N=8$ particles when viewed 
with respect to the global $N$. Next, we demonstrate the dependence 
of the CFs $n_{(ijk)}$ and $\tilde{n}_{(ijk)}$ on the HC diameter 
$a_c$, where an important role for the interference between the 
lattice wells is revealed. Similarly, the global and local SFFs, 
$\rho_s/\rho$ and $(\rho_s/\rho)_{(ijk)}$, respectively, are 
explored as functions of $a_c$.

\hs From a global perspective, we chiefly find a counter-intuitive,
opposing behavior for the global CF and SFF as functions of 
$k\,=\,\pi/d$. Peculiarly, the SFF drops with increasing $k$ 
while the CF rises. Additionally, we seek a MI state by going 
to larger interactions or OL depth, and show that our systems 
display a coexistence of SF and vacuum MI regimes. 

\hs Finally, we also present results for the momentum distributions 
for some of our systems. The latter reveal a diffractive structure 
which persists deep into the MI regime \cite{Spielman:2007} suggesting 
that superfludity \cite{Spielman:2007} in a CHOCL is found even 
in the extremely repulsive regime.

\begin{figure}[t!]
\includegraphics[width=8.5cm,bb=158 500 446 728,clip]{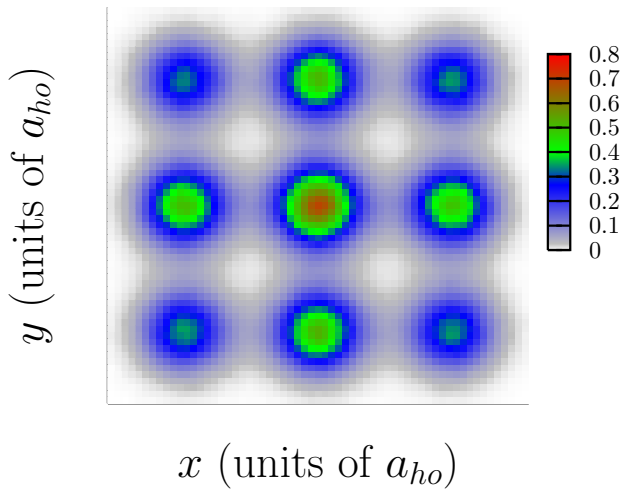}
\caption{Density map for a system of $N=8$ bosons,
$V_0=10$, and $k=1.0\pi$ in the same confining geometry
as Fig.\ft\ref{fig:plot.n000_vs_ac.N8.B10.several.k.stack}
below. This map is obtained by integrating the three-dimensional
density along the $z-$axis.}
\label{fig:plot.od.vmc.N8.B10.k1.0.L27.ac0.14.mapview}
\end{figure}

\begin{figure}[t!]
\includegraphics[width=8.5cm,bb=155 520 502 772,clip]{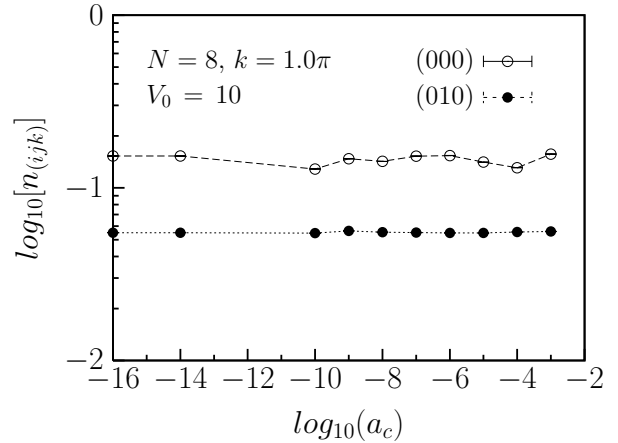}
\caption{CF $n_{(ijk)}\,=\,N_{(ijk)}/N$ at the central cell 
(open circles) and $(010)$ (solid circles) as a function of 
$a_c$ in the extremely dilute regime. The system is a HC Bose 
gas of $N=8$ particles, $V_0=10$, and $k=1.0\pi$ in the same 
confining geometry as 
Fig.\ft\ref{fig:plot.n000_vs_ac.N8.B10.several.k.stack} below.
A log-log scale is used in order to provide a clearer view of the 
details. The $a_c$, $k$ and $V_0$ are in units of $a_{ho}$, 
$a_{ho}^{-1}$ and \hw, respectively, where \aho.}
\label{fig:plot.VMC.DILUTE.GAS.Condensate.fractions.n0.vs.ac.N8.k1.0.L27}
\end{figure}

\subsection{Numerics}

\hs For the CFs at the lattice sites $(ijk)$, the configurations 
of previous VMC simulations are used \cite{Sakhel:2010}. The local 
CF is computed as in Sec.\ref{sec:density-matrix}, whereas the 
global one by the additional Eq.(\ref{eq:total-condensate-fraction}). 
The global SFF is computed using Eq.(\ref{eq:VPI-superfluid-fraction}) 
and the local SFF by an additional application of the boundary conditions
(\ref{eq:SFinsideaCell}). We use $N=8$ particles with $V_0=10$, and
$N=40$ with $V_0=10$ and once with $V_0=20$. The total number of 
VPI time slices used is $M=120$ with a time step $\tau\sim O(10^{-3})$. 
The number of MC steps and MC blocks is $\sim O(10^3)$. 

\hs In order to justify the method outlined in Sec.\ref{sec:method},
necessiating the presence of spherically symmetric clouds centered
at each lattice site, we present in 
Fig.\ft\ref{fig:plot.od.vmc.N8.B10.k1.0.L27.ac0.14.mapview} a map 
for the integrated VMC density along the $z-$axis 
$\langle n_{2D}(x,y)\rangle$ for the system of 
Fig.\ft\ref{fig:plot.n000_vs_ac.N8.B10.several.k.stack} below. One 
can see clearly that the distributions of the atoms about the 
lattice sites are spherical at the edges of the trap. Thus it can 
be safely stated, that for a few bosons spherical symmetry is 
displayed by the lattice clouds, in spite of the presence of an 
external harmonic trap. Yet by increasing the number of particles, 
it is anticipated that the clouds at the edges of the trap will 
be deformed from their spherically symmetric form, as the atoms 
are pushed away from the trap center towards the edges of the
trap by their mutual HC repulsion. This is especially the case,
when $a_c$ becomes large. Consequently, in the absence of 
spherical symmetry, the method of Sec.\ref{sec:method} cannot be used 
anymore, except for the clouds which may nevertheless preserve their
spherical form. If one uses a large CHOCL with, say, 
$N_L\,=\,5\times 5\times 5$ lattice sites, one must consider a 
weaker harmonic trap than the one used here in order to maintain 
spherically symmetric clouds at the edges of the trap.

\subsection{Condensate fraction and interference effects}

\hs In this section, we explore the local and global BEC
in our CHOCL systems. Only five lattice sites, (000), (001),
(010), (011), and (111), which present the whole CHOCL system
are considered. The role of interference between the lattice
wells is revealed by a comparison between $n_{(ijk)}$ and
$\tilde{n}_(ijk)$.

\begin{figure}[t!]
\includegraphics[width=7.5cm,bb = 165 288 436 772,clip]{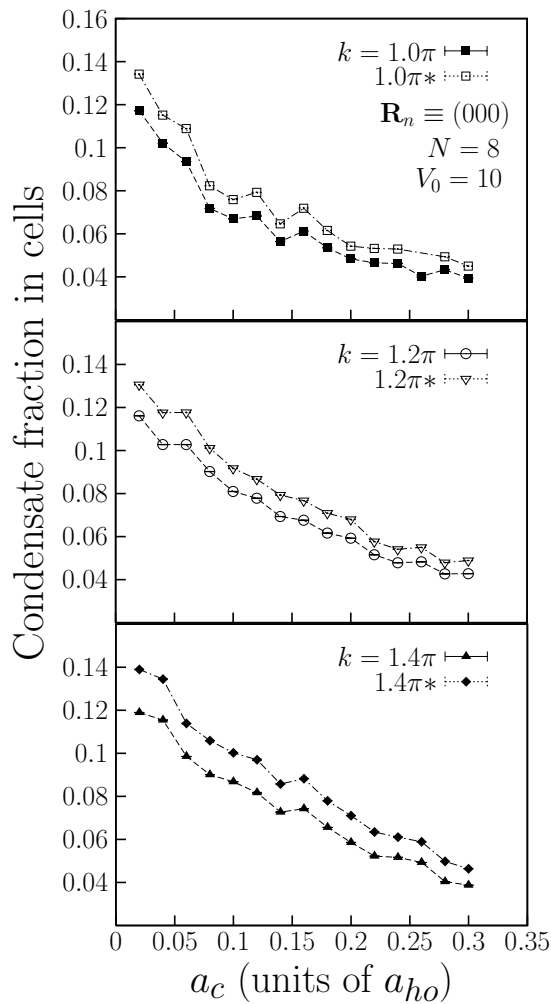}
\caption{CFs $n_{(000)}$ and $\tilde{n}_{(000)}$ 
at lattice site $\mathbf{R}_n\equiv(000)$ versus the HC 
diameter $a_c$ for systems with and without interference effects, 
respectively, and at the various $k-$values indicated. The systems 
are the same as in Ref.\cite{Sakhel:2010}: $N=8$ HC bosons 
confined in a cubic OL of $N_L=3\times 3\times 3$ 
lattice sites plus external harmonic confinement. The height
of the OL barrier is $V_0=10$. The $\tilde{n}_{(000)}$ 
values are labelled by ($*$) in each legend. Solid squares: $k=1.0\pi$; 
open squares $k=1.0\pi *$; open circles: $1.2\pi$; open triangles 
$1.2\pi *$; solid triangles: $1.4\pi$; diamonds $1.4\pi *$. The $k$ 
and $V_0$ are in units of $a_{ho}^{-1}$ and \hw, where \aho.}
\label{fig:plot.n000_vs_ac.N8.B10.several.k.stack}
\end{figure}

\subsubsection{Local CF in ideal BEC}

\hs We begin with an investigation of the ideal Bose gas 
for $a_c$ values ranging from $10^{-16}$ to $10^{-3}$ and 
show that in this regime the local CF is almost stable. The 
$a_c$ values range from an extremely dilute to dilute regime.
Fig.\ft\ref{fig:plot.VMC.DILUTE.GAS.Condensate.fractions.n0.vs.ac.N8.k1.0.L27}
displays the local CF at the central cell (open red circles) 
and $(010)$ (solid blue circles). We get $n_{(000)}\sim 15\%$ and 
$n_{(010)}\sim 5.5\%$ of the total $N$, which in themselves are
small CFs. It should be emphasized that the optimized VMC parameters 
of the trial wave function (\ref{eq:trial-wave-function}) are 
exactly the same for all $a_c$ in the dilute regime. This is because 
in the dilute regime, the wave function of the system 
[Eq.(\ref{eq:trial-wave-function})] is essentially a product of 
the Wannier sums Eq.(\ref{eq:Wannier-like-function}) with little 
effect coming from the Jastrow function.

\begin{figure}[t!]
\includegraphics[width=8.5cm,bb=170 542 507 790,clip]{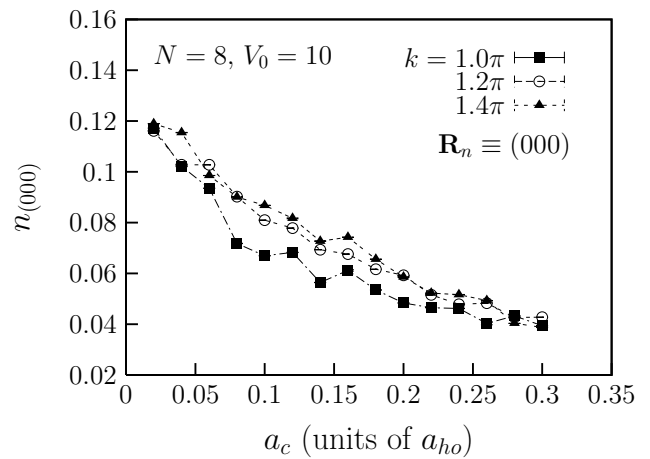}
\caption{Same systems as in 
Fig.\ft\ref{fig:plot.n000_vs_ac.N8.B10.several.k.stack}; but
without interference data. The $k$ and $V_0$ are in units of
$a_{ho}^{-1}$ and \hw, respectively.}
\label{fig:plot.n000_vs_ac.N8.B10.several.k}
\end{figure}

\begin{figure}[t!]
\includegraphics[width=8.0cm,bb = 165 202 436 768,clip]{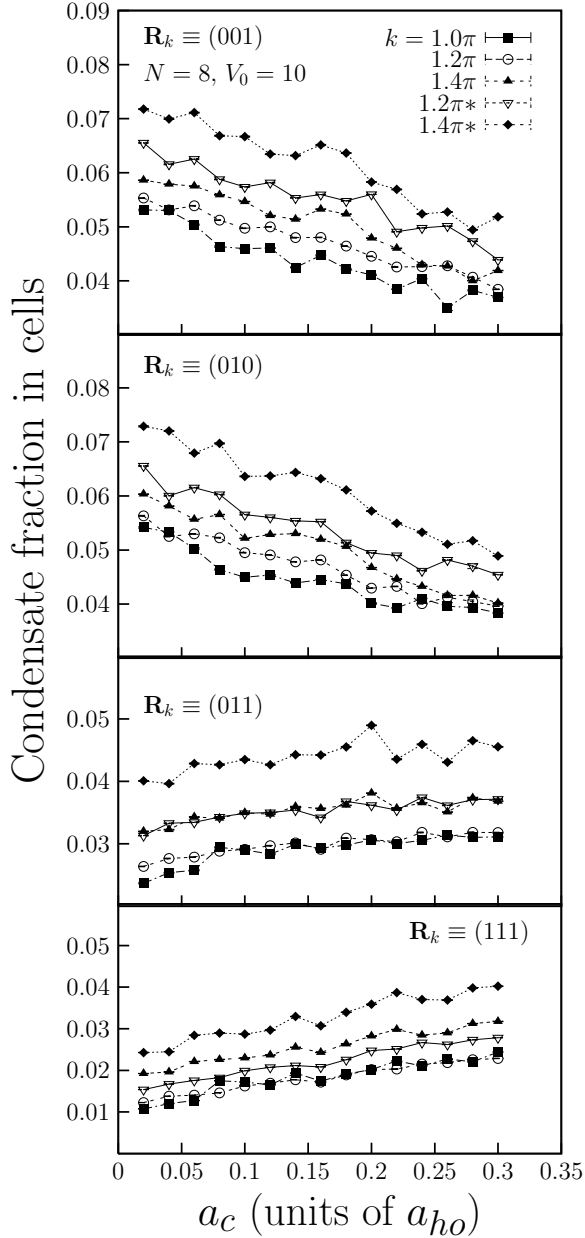}
\caption{As in Fig.\ft\ref{fig:plot.n000_vs_ac.N8.B10.several.k.stack}; 
but from top to botton for lattice sites $\mathbf{R}_n\equiv(001)$,
(010), (011), and (111). The $k$ and $V_0$ are in units of $a_{ho}^{-1}$ 
and \hw, where \aho.}
\label{fig:plot.nntilde_vs_ac.N8.B10.several.k.stack}
\end{figure}

\begin{figure}[t!]
\includegraphics[width=8.5cm,bb = 170 544 508 793,clip]{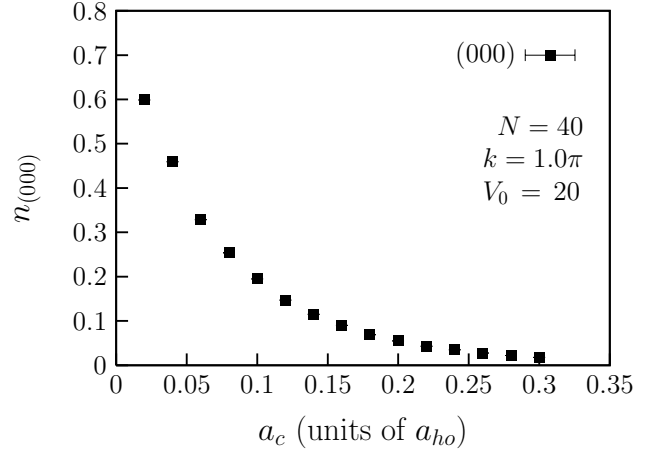}
\caption{CF $n_{(000)}$ versus the 
HC diameter $a_c$ for a system of $N=40$, $V_0=20$, 
and $k=1.0\pi$ in the same trapping geometry of 
Fig.\ft\ref{fig:plot.n000_vs_ac.N8.B10.several.k.stack}. The 
$k$ and $V_0$ are in units of $a_{ho}^{-1}$ and \hw, respectively,
where \aho.}
\label{fig:plot.n0_vs_ac.N40.B10.several.Rn}
\end{figure}

\subsubsection{Local CF in the interacting BEC}

\hs Fig.\ft\ref{fig:plot.n000_vs_ac.N8.B10.several.k.stack} displays 
the CFs $n_{(000)}$ and $\tilde{n}_{(000)}$ (starry labels) in the 
central well $\mathbf{R}_n\equiv (000)$ versus $a_c$ at the indicated 
values of $k$. From the top frame to the bottom frame we display for 
$n_{(000)}$: (solid squares) $k=1.0\pi$; (open circles) $1.2\pi$; 
(solid up triangles) $1.4\pi$. The systems are the same as in 
Ref.\cite{Sakhel:2010}. For $\tilde{n}_{(000)}$ we display: (open squares) 
$k\,=\,1.0\pi*$, (open down triangles) $1.2\pi*$, and (diamonds) 
$1.4\pi*$.
Fig.\ft\ref{fig:plot.n000_vs_ac.N8.B10.several.k} is the same as
Fig.\ft\ref{fig:plot.n000_vs_ac.N8.B10.several.k.stack}, but
without interference data. The goal of 
Fig.\ft\ref{fig:plot.n000_vs_ac.N8.B10.several.k} is to give further
manifestation to the dependence of $n_{(000)}$ on $k$. One can see 
that for all values of $k$ in the latter figures, $n_{(000)}$ and 
$\tilde{n}_{(000)}$ decrease with an increase in $a_c$. Further, 
the $\tilde{n}_{(000)}$ lies higher than $n_{(000)}$ for all $k$, 
indicating that interference between the lattice wells boosts the 
CF in each well.
In Fig.\ft\ref{fig:plot.n000_vs_ac.N8.B10.several.k}, the values 
of $n_{(000)}$ seem to rise with increasing $k$ in the range 
$0.05 \le a_c \le 0.25$, although for $k=1.2\pi$ and $1.4\pi$ the 
values are very close to each other.

\hs Fig.\ft\ref{fig:plot.nntilde_vs_ac.N8.B10.several.k.stack} is 
the same as 
Fig.\ft\ref{fig:plot.n000_vs_ac.N8.B10.several.k.stack}, but for
$\mathbf{R}_n\equiv (001)$, (010), (011), and (111), ordered 
respectively from the top frame to the bottom frame. The same
labels are used as in 
Fig.\ft\ref{fig:plot.n000_vs_ac.N8.B10.several.k.stack}. However,
whereas $n$ and $\tilde{n}$ decrease with increasing $a_c$ for
(001) and (010), they surprisingly increase for (111) and (011).
Some condensate has tunneled away from the sites 
$\mathbf{R}_n\equiv (000)$, (001), and (010) towards (011) and
(111) as a result of an increase in the mutual repulsion of the
bosons which drives them away from the center of the trap. A 
similar expulsion of the BEC towards the edges of the trap has
also been encountered in the simple harmonic trap without an OL
\cite{DuBois:01,Sakhel:2008}. This phenomenon corresponds also 
to a rise in the occupancy of the lattice wells at the corners 
of the CHOCL with increasing $a_c$, as demonstrated in 
Sec.\ref{sec:canwgMI} later on. Otherwise, 
Fig.\ft\ref{fig:plot.nntilde_vs_ac.N8.B10.several.k.stack} 
displays the same features as 
Fig.\ft\ref{fig:plot.n000_vs_ac.N8.B10.several.k.stack} regarding
the values of $k$ and the interference effects. The rise of 
$n_{(ijk)}$ with increasing $k$ is even more pronounced in 
Fig.\ft\ref{fig:plot.nntilde_vs_ac.N8.B10.several.k.stack}. 
One can explain this by approximating each lattice well by 
a HO trap, whose trapping frequency $\omega_k$ rises with 
increasing lattice wave vector $k\,=\,\pi/d$. As a result of 
increasing $\omega_k$, the level spacing $\sim\hbar\omega_k$ 
becomes larger, thus increasing the energy needed to excite 
atoms out of the ground state HO level. Consequently, more 
particles find themselves forced to condense into the lowest 
``harmonic oscillator" level. (We did not include in 
Fig.\ft\ref{fig:plot.nntilde_vs_ac.N8.B10.several.k.stack} $\tilde{n}$
for $k=1.0\pi*$ in order not to clutter the figures.)

\hs Another explaination is as follows. Going back to Fig.\ft15 
of Ref.\cite{Sakhel:2010}, one can see that the occupancy of 
the central well $(000)$ and the first neighbor (00 $-$1) decreases 
with increasing $k$. [Note that (00 $-$1) belongs to the same 
lattice site family of $(010)$ and $(001)$.] Consequently, the 
reduction in the occupancy reduces the density in these lattice 
wells, and henceforth the onsite interactions between the bosons. 
This in turn reduces the depletion effect of the HCs causing a rise 
in the CFs with increasing $k$ as observed in 
Figs.\ft\ref{fig:plot.n000_vs_ac.N8.B10.several.k.stack}$-$\ref{fig:plot.nntilde_vs_ac.N8.B10.several.k.stack}.
It is further noted that $n_{(001)}$ is changing with $a_c$ at a lower 
rate than $n_{(000)}$. E.g, $n_{(001)}$ for $k=1.0\pi$ (solid 
squares) in Fig.\ft\ref{fig:plot.nntilde_vs_ac.N8.B10.several.k.stack} 
declines from $\sim0.055$ down to $\sim0.04$ over the whole range
of $a_c$ in the figure, whereas $n_{(000)}$ for the same $k$ (see 
Fig.\ft\ref{fig:plot.n000_vs_ac.N8.B10.several.k}) decreases from 
$\sim0.12$ to $\sim0.04$. The same feature is observed for 
$\mathbf{R}_n\equiv (010)$ when compared to (000). In contrast, 
$n_{(111)}$ and $n_{(011)}$ increase by $\sim0.01$ in 
Fig.\ft\ref{fig:plot.nntilde_vs_ac.N8.B10.several.k.stack}.
Hence, the rate of change of the $n_{(ijk)}$ with $a_c$, whether
$n_{(ijk)}$ is rising or declining, becomes smaller as the 
lattice well approaches the edges of the trap. This is attributed 
to the fact that the energy cost required for the bosons to tunnel 
through (or between) two neighboring CHOCL wells increases as the 
lattice wells approach the edges of the background harmonic trap. 
This is because each lattice well is superimposed on the harmonic 
trap. Therefore the potential energy minimum of each lattice 
well at position $\mathbf{R}_n$ from the trap center, is equal to 
the HO energy $(1/2) |\mathbf{R}_n|^2$ and rises with increasing 
$|\mathbf{R}_n|$. In addition, we present in 
Fig.\ft\ref{fig:plot.n0_vs_ac.N40.B10.several.Rn} the behavior 
of the CF $n_{(000)}$ for $N=40$ particles, $V_0=20$, and $k=\pi$. 
Again, the BEC gets depleted with increasing $a_c$, except that 
the CFs are substantially higher than for $N=8$. 

\hs It must be emphasized, that the values of the CFs 
in Figs.\ft\ref{fig:plot.n000_vs_ac.N8.B10.several.k.stack}$-$\ref{fig:plot.nntilde_vs_ac.N8.B10.several.k.stack}
fluctuate 
with $a_c$ due to the small number of particles, $N=8$, used
which causes real physical fluctuations to be large \cite{Mullin:2011}. 
Nevertheless, the trend in the CFs clearly indicates depletion 
with increasing $a_c$ for some lattice cells and a condensate 
buildup in other cells. It is further noted, that in general 
the CFs for $N=8$ are significantly lower than 1, bringing this
result in line with the earlier findings of Sun \ea\ \cite{Sun:2009}.
For $N=40$, the data varies smoothly with $a_c$ and the statistics
are good, demonstrating that a higher $N$ leads to lower condensate 
number fluctuations inside the OL.

\begin{figure}[b!]
\includegraphics[width=8.5cm,bb=173 544 507 790,clip]{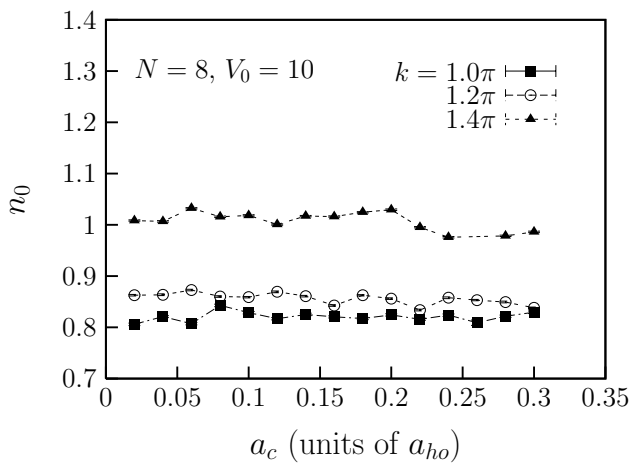}
\caption{Global CF $n_0$ as estimated by 
Eq.(\ref{eq:total-condensate-fraction}) for the systems
in Fig.\ft\ref{fig:plot.n000_vs_ac.N8.B10.several.k.stack} 
at $k=1.0\pi$ (solid squares); $k=1.2\pi$ (open circles), 
and $k=1.4\pi$ (solid triangles). The $k$ and $V_0$ are in 
units of $a_{ho}^{-1}$ and \hw, respectively, where \aho} 
\label{fig:plot.total.n0.vs.ac.N8.B10.several.k}
\end{figure}

\subsubsection{Global CF $n_0$}

\hs Using the previous results of 
Figs.\ft\ref{fig:plot.n000_vs_ac.N8.B10.several.k.stack}$-$
\ref{fig:plot.nntilde_vs_ac.N8.B10.several.k.stack} for the 
noninterfering case only, we can estimate the global CF $n_0$ 
using Eq.(\ref{eq:total-condensate-fraction}). The goal is to 
distinguish between condensate local and global behavior. 
Fig.\ft\ref{fig:plot.total.n0.vs.ac.N8.B10.several.k} displays 
$n_0$ for the systems in the latter figures at $k=1.0\pi$ 
(solid squares), $1.2\pi$ (open circles), and $1.4\pi$ (solid 
triangles). It is noted, that $n_0$ is practically constant 
and not sensitive to the changes in the repulsive forces via 
$a_c$. This is a peculiar result, suggesting that the CF is 
globally conserved. In contrast, the decline of $n_{(ijk)}$ 
with increasing $a_c$ at the lattice sites (000), (001), and 
(010) in Figs.\ft\ref{fig:plot.n000_vs_ac.N8.B10.several.k.stack}$-$
\ref{fig:plot.nntilde_vs_ac.N8.B10.several.k.stack} is a local
effect and was compensated by a corresponding rise at (011) 
and (111). The $n_0$ for $k=1.0\pi$ is $\sim 80\%$ for all
values of $a_c$, whereas for $k=1.2\pi$ and $1.4\pi$, 
$n_0\sim 85\%$ and $\sim 100\%$, respectively, for all $a_c$.
The values of $n_0$ are high because for $N=8$ particles the
systems are in the dilute regime and remain in this state even
up to very large HC repulsion. Further, the global CF is 
boosted with the rise of $k$ (the local-confinement strength
$\omega_k\propto k\,=\pi/d$) in each CHOCL well. The OL 
introduces a local depletion effect at lower $k<1.4\pi$ when 
there is almost no role for the HC repulsion in the global 
depletion. An important finding, then, is that the BEC in a
CHOCL with a few bosons is chiefly depleted by the OL. 
Therefore, whereas the global CF $n_0$ remains constant as $a_c$ 
is changed, the local CF is redistributed at the various lattice 
sites. In contrast, Ramanan \ea\ \cite{Ramanan:2009} found that 
for a 1D Bose gas trapped by an OL plus a weak external harmonic 
trap, the total CF is depleted substantially with an increase 
of the onsite repulsive energy $U$. Using the BHM, van Oosten
\ea\ \cite{vanOosten:2001} demonstrated that the CF in a 2D
and 3D OL decreases with increasing parameter $U/t$, where $U$
is the contact interactions strength and $t$ the hopping
amplitude.

\subsection{Superfluid fraction}\label{sec:superfluid-fraction}

\hs In this section, we explore the SFF in our CHOCL systems 
both as a global and as a local quantity in individual lattice 
wells. We consider systems with $N=8$ and $N=40$ bosons in the 
same confining geometry as in 
Fig.\ft\ref{fig:plot.n000_vs_ac.N8.B10.several.k.stack} with 
the same $k-$values and $V_0=10$. The SFF is computed globally 
using Eq.(\ref{eq:VPI-superfluid-fraction}), and locally for 
$a_c\le 0.3$ by Eq.(\ref{eq:SFinsideaCell-semi-global-view}) and
an additional application of the boundary conditions 
(\ref{eq:SFinsideaCell}). For the larger $a_c\sim O(1)$, the 
SFF was computed with respect to the {\it total} $N$ as outlined
later below. We did not compute the SFF for the systems with 
$N=8$ particles for the range of $a_c\le 0.3$ considered so 
far because of the high statistical fluctuations in the 
values of the VPI MC block averages. However, at extreme 
values of $a_c$ beyond 1.00, these fluctuations are much lower, 
and therefore it was reasonable to report the SFF for $N=8$ 
as in Sec.\ref{sec:canwgMI} below.

\subsubsection{Global SFF $\rho_s/\rho$}

\begin{figure}[t!]
\includegraphics[width=8.5cm,bb = 175 544 508 794,clip]{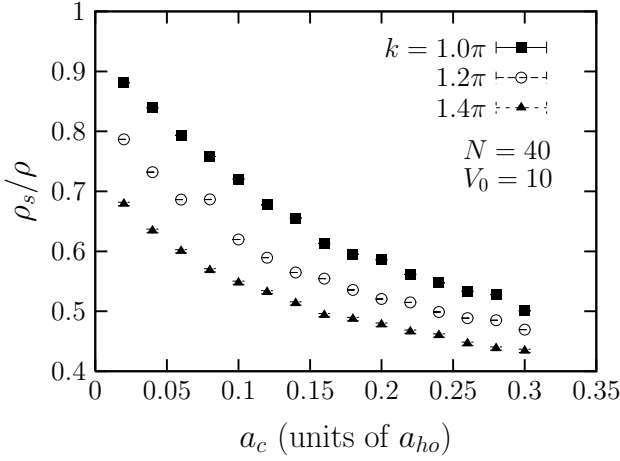}
\caption{Global SFF $\rho_s/\rho$ versus the HC diameter $a_c$ 
as computed by the diffusion formula of Pollock and Ceperely 
\cite{Pollock:1987} given here by Eq.(\ref{eq:VPI-superfluid-fraction}). 
The system considered is $N=40$ HC bosons in the same trapping 
geometry of
Fig.\ref{fig:plot.n000_vs_ac.N8.B10.several.k.stack} and the 
indicated values of $k$. The same labels are used as in 
Fig.\ft\ref{fig:plot.nntilde_vs_ac.N8.B10.several.k.stack}. The 
$k$ and $V_0$ are in units of $a_{ho}^{-1}$ and \hw, respectively, 
where \aho.} 
\label{fig:plot.vpi.rho_sf.vs.ac.diffusion.N40.B10.severalk}
\end{figure}

\hs Fig.\ft\ref{fig:plot.vpi.rho_sf.vs.ac.diffusion.N40.B10.severalk}
displays the global SFF, $\rho_s/\rho$, versus $a_c$ for the 
latter systems at the indicated values of $k$. The labelling is as in 
Fig.\ft\ref{fig:plot.total.n0.vs.ac.N8.B10.several.k}. One can see 
that the global SF is depleted with a rise in $a_c$. Further, the values 
of $\rho_s/\rho$ decrease with increasing $k$, and one can conclude 
that a higher $k$ (lower lattice spacing) enhances the depletion of 
the global SF in an OL. This is our chief result and is counter-intuitive 
to the rise of the global CF with increasing $k$ as in 
Fig.\ft\ref{fig:plot.total.n0.vs.ac.N8.B10.several.k}. 
Retrospectively, this could also be explained as before 
\cite{Sakhel:2010}: a smaller lattice spacing increases the localization
of the bosons inside each lattice cell and causes a reduction in 
their superflow. A reduction in superflow means a smaller number 
of particles possessing enough kinetic energy for tunneling from 
one well to another. Further note that remarkably even at very 
strong repulsion between the bosons, superfluidity is still present
($\rho_s/\rho\stackrel{>}{\sim} 40\%$ for $a_c\,=\,0.3$.) In 
contrast to the global BEC in 
Fig.\ft\ref{fig:plot.total.n0.vs.ac.N8.B10.several.k}, $a_c$ plays
a chief role in the global depletion of the SF in 
Fig.\ft\ref{fig:plot.vpi.rho_sf.vs.ac.diffusion.N40.B10.severalk}.

\begin{figure}[t!]
\includegraphics[width=8.5cm,bb = 162 280 439 771,clip]{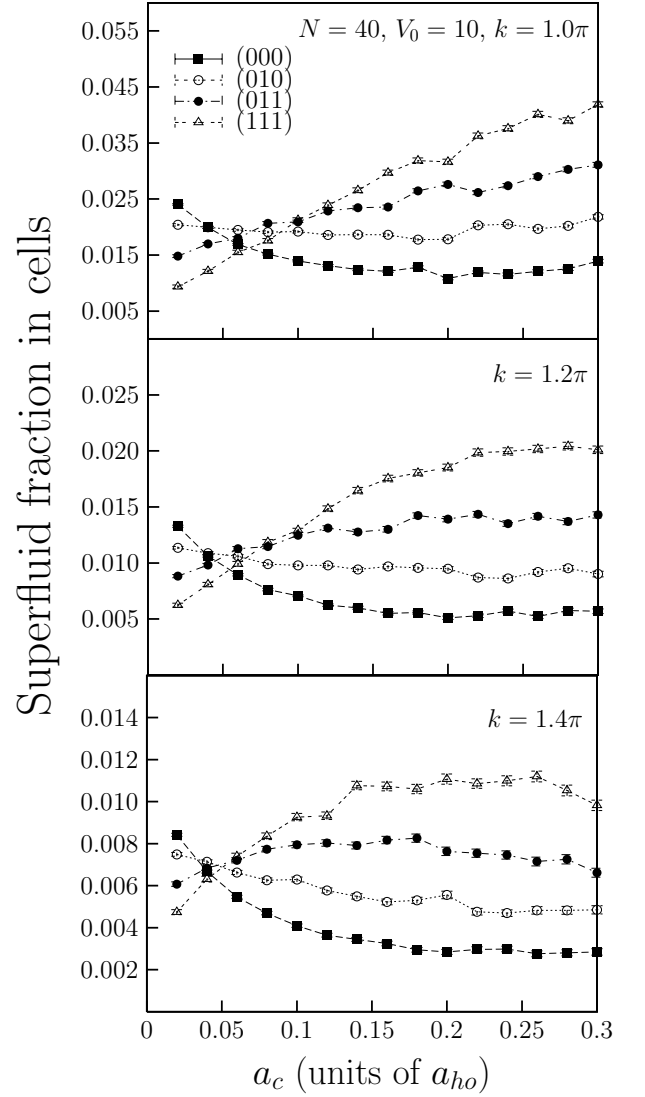}
\caption{As in Fig.\ft\ref{fig:plot.vpi.rho_sf.vs.ac.diffusion.N40.B10.severalk};
but for the local SFF $(\rho_s/\rho)_{(ijk)}$ at the indicated 
values of $k$. The cell boundaries are given by Eq.(\ref{eq:SFinsideaCell}). 
The $(\rho_s/\rho)_{(ijk)}$ is computed via
Eq.(\ref{eq:SFinsideaCell-semi-global-view}) by using the number 
of particles $N_{(ijk)}$ inside each cell instead of the total number 
of particles $N$. Solid squares: $(ijk)\equiv(000)$; open cirlces: 
(010); solid circles: (011); open triangles: (111). The $k$ and $V_0$ 
are in units of $a_{ho}^{-1}$ and \hw, respectively, where \aho.}
\label{fig:plot.VPI.superfuid.fr.N40B10.several.k.cells}
\end{figure}

\subsubsection{Local SFF $(\rho_s/\rho)_{(ijk)}$}\label{sec:LSFF}

\hs As for the SFF in each CHOCL cell, $(\rho_s/\rho)_{(ijk)}$, it 
reveals a different behavior as compared to its global character in 
Fig.\ft\ref{fig:plot.vpi.rho_sf.vs.ac.diffusion.N40.B10.severalk}.
Fig.\ft\ref{fig:plot.VPI.superfuid.fr.N40B10.several.k.cells} 
displays $(\rho_s/\rho)_{(ijk)}$ vs. $a_c$ for the same systems 
of Fig.\ft\ref{fig:plot.vpi.rho_sf.vs.ac.diffusion.N40.B10.severalk},
where for the top frame to the bottom frame: $k\,=\,1.0\pi$, $1.2\pi$, 
and $1.4\pi$, respectively. Solid squares: $(ijk)\equiv(000)$; open 
circles: (010); solid circles: (011); open triangles: (111). Inside 
each cell, the local $(\rho_s/\rho)_{(ijk)}$ is computed using 
Eq.(\ref{eq:SFinsideaCell-semi-global-view}) plus the condition 
(\ref{eq:SFinsideaCell}). 

\hs The same features are observed in all three frames: 
$(\rho_s/\rho)_{(011)}$ and $(\rho_s/\rho)_{(111)}$ rise with increasing 
$a_c$. The $(\rho_s/\rho)_{(010)}$ displays an initial weak decline, but 
then it stabilizes somewhat after $a_c=0.2$. For the center (000) the 
decline is more pronounced than in (010) up to $a_c=0.16$, after which 
it stabilizes somewhat. In fact, the rise of $(\rho_s/\rho)_{(011)}$ 
and $(\rho_s/\rho)_{(111)}$ with increasing $a_c$ conforms to the rise 
of the CFs $n_{(011)}$ and $n_{(111)}$ with $a_c$ in 
Fig.\ft\ref{fig:plot.nntilde_vs_ac.N8.B10.several.k.stack} (although for
a different number of particles), whereas the decline in (000) corresponds 
to that of $n_{(000)}$ in Fig.\ft\ref{fig:plot.n000_vs_ac.N8.B10.several.k}. 
It can therefore be argued, that locally the behavior of the SF
is isomorphic to that of the condensate. Although $N$ is different, 
it is important to emphasize that the local SFFs in 
Fig.\ft\ref{fig:plot.VPI.superfuid.fr.N40B10.several.k.cells} are 
of the same order of magnitude as the local CFs in 
Fig.\ft\ref{fig:plot.nntilde_vs_ac.N8.B10.several.k.stack}. The 
reduction in $(\rho_s/\rho)_{(000)}$ is in line with the previous 
finding \cite{Sakhel:2010}, that a rise in $a_c$ reduces the 
single-particle tunneling amplitude $J$ between, e.g., the central 
(000) and first neighbor (010) cells. The superflow is therefore 
suppressed between those lattice wells because of an increased 
localization at the larger $a_c$.

\hs By inspecting  
Fig.\ft\ref{fig:plot.VPI.superfuid.fr.N40B10.several.k.cells}, one 
notes again a similar counter-intuitive feature as for the global 
values; that for each cell the local SFF drops with increasing $k$ 
contrary to the corresponding local CF in
Fig.\ft\ref{fig:plot.nntilde_vs_ac.N8.B10.several.k.stack}. 
Another peculiar feature in 
Fig.\ft\ref{fig:plot.VPI.superfuid.fr.N40B10.several.k.cells}, is 
that all four curves in each frame intersect almost at the same 
value of $a_c$ where there is almost an equal distribution of 
the SF at all lattice sites. The SF migrates from the cells near 
the center of the trap towards the corners of the CHOCL with the 
increase of $a_c$. This feature is also elaborated in 
Sec.\ref{sec:canwgMI} below, where systems with extremely repulsive 
HCs are investigated.

\hs The effects of an external harmonic trap are as such to introduce
a new behavior for the superflow of bosons between the wells of an
OL. As the repulsive interactions become stronger, the SF is 
``expelled" towards the edges of the trap analogous to the
condensate in a simple harmonic trap \cite{DuBois:01,Sakhel:2008}.
However, in a CHOCL parts of the condensate still remain situated
in the lattice wells close to the center of the trap.

\subsection{Energies}

\begin{figure}[t!]
\includegraphics[width=8.5cm,bb = 175 544 508 794,clip]{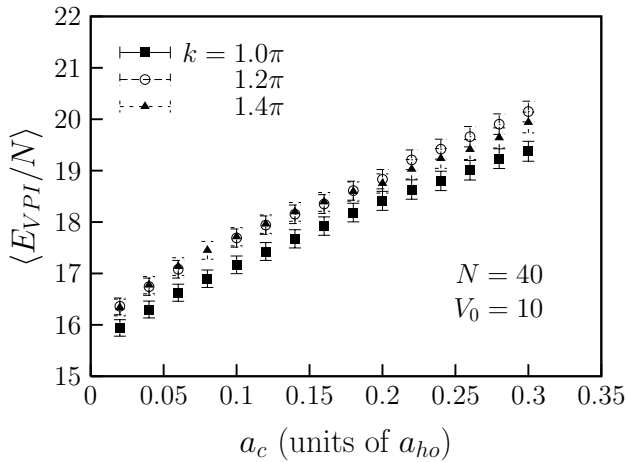}
\caption{Average VPI energy per particle, $\langle E_{VPI}/N\rangle$, 
versus the HC diameter $a_c$ for the same systems of 
Fig.\ft\ref{fig:plot.vpi.rho_sf.vs.ac.diffusion.N40.B10.severalk}.
The $k$ is in units of $a_{ho}^{-1}$ where \aho. The 
$\langle E_{VPI}/N\rangle$ and $V_0$ are in units of \hw.} 
\label{fig:plot.vpi.energy.vs.ac.N40.B10.severalk}
\end{figure}

\hs Fig.\ft\ref{fig:plot.vpi.energy.vs.ac.N40.B10.severalk} displays
the behavior of the average VPI energy per particle, $\langle E_{VPI}\rangle/N$,
as a function of the HC diameter $a_c$ for the same systems 
in Fig.\ft\ref{fig:plot.vpi.rho_sf.vs.ac.diffusion.N40.B10.severalk}.
Whereas $\langle E_{VPI}/N\rangle$ rises with increasing $a_c$, 
$\rho_s/\rho$ decreases 
(see Fig.\ft\ref{fig:plot.vpi.rho_sf.vs.ac.diffusion.N40.B10.severalk}).
The rise in the energy is chiefly due to the rise in the average onsite
interaction energy $\langle U_{(pqr)}\rangle$ with $a_c$, as found earlier
in Ref.\cite{Sakhel:2010}. Therefore, the later rise overwhelms the 
drop in kinetic energy, i.e. mobility, in favor of an increase in boson
localization.

\subsection{Can we get a {\it pure} Mott insulator?}\label{sec:canwgMI}

\hs In this section, we show that for a few bosons in a CHOCL 
with a limited number of lattice sites, MI regimes can be 
achieved by increasing the HC diameter of the bosons to large 
values. It is found that only a part of the system will be in the
MI state where a number of CHOCL wells display the absence of
superfluidity. We show that our systems are different than the 
ones usually treated by the BHM for a much larger number of 
particles $N$. For example, Hen and Rigol \cite{Hen:2010b} 
explored the phase diagram of a HC BHM on a checkerboard superlattice. 
According to Hen and Rigol \cite{Hen:2010a,Hen:2010b,Rigol:2009}, 
bosons inside a CHOCL display coexisting SF and MI phases. 
Mixed SF-MI regimes have also been experimentally reported by 
Spielman \ea\ \cite{Spielman:2007,Spielman:2008} in 2D atomic 
gases confined in harmonic OL potentials. In the work of Jaksch 
\ea\ \cite{Jaksch:1998}, a checkerboard SF-MI phase was observed 
in an OL. Their most important finding was that a BHM could be 
realized by the dynamics of bosons in an OL. But this was found 
for a much larger number of particles and lattice sites than ours, 
using a discrete space approach, whereas our investigation is 
restricted to a few bosons and lattice sites in continuous space. 
 
\hs Hen and Rigol \cite{Hen:2010a} explored the ground state 
properties of HC bosons in 2D and 3D OLs confined by an external 
harmonic trap. It was demonstrated that a MI state usually displays 
a flat density along several lattice sites, as in their Figs.\ft 7 
and 14, where the density is displayed as a function of the distance 
from the center of the trap. However, a limited number of lattice 
sites as in the present work here does not furnish the ground 
for obtaining a flat density profile.

\hs In the investigations of Hen and Rigol 
\cite{Hen:2010a,Hen:2010b,Hen:2009} and Jaksch \ea\ \cite{Jaksch:1998} 
the interactions are described by contact potentials and have no 
range ($a_c$) as in our case. Thus, the BHM cannot be used here 
to make predictions about a possible SF-MI transition. In fact, 
a large increase in $a_c$ causes $-$as shown next$-$ the bosons 
to be driven out of the center of the trap towards the corners of 
the CHOCL, instead of making them prefer to occupy sites individually 
as in the BHM. This is because when the volume of each HS boson is 
increased, they have no other way but to increase their minimum 
separation as they are not able to approach each other by a distance
less than $a_c$, prescribed by the Jastrow function (\ref{eq:Jastrow}).
Therefore, a homogeneous distribution cannot be achieved. Hence, 
increasing $a_c$ largely in our systems does not yield a pure MI 
state throughout the whole lattice wells. We are thus very much 
inclined to say, given the above information, that our systems will 
retain an SF even in the strongly interacting regime. 

\hs In the BHM, on the other hand, a large increase in the 
repulsive contact potential makes it energetically more costly 
to hop from one lattice well to another, but it does not increase 
the volume of the HS bosons. A homogeneous distribution is thus 
naturally possible in the BHM as the bosons seek the lowest 
energetic configuration in the OL. 

\hs Further, in contrast to Hen and Rigol \cite{Hen:2010a}, the 
vacuum surrounding our CHOCL does not contain empty lattice 
sites, and therefore cannot be associated with an empty MI state. 
The empty MI regime was mentioned earlier \cite{Hen:2010a}. 
However, inside the CHOCL it will be demonstrated that vacuum MI 
states are possible.

\hs In order to check for the presence of a SF phase in some parts 
of our systems for large $a_c$, one could compute $U/(z J)$ 
between two neighboring lattice sites. This should help us shed
more light on the state of the current CHOCL systems, because it 
is hard to distinguish between a SF and MI by only looking at the 
densities. The identification of MI domains only by means of the 
density was found to be inaccurate, as outlined earlier by Rigol 
\ea\ \cite{Rigol:2009} (and references therein). Earlier, it was 
found \cite{Sakhel:2010} that for $a_c$ up to 0.3 our systems
were still in the SF phase and that $U/(z J)$ began to stabilize
for large $a_c$. For the tunneling from the central cell (000)
to the first nearest neighbor (00 $-1$), our calculations yielded
a $U/(z J) \ll 5.814$ (see Fig.6 in Ref.\cite{Sakhel:2010}). However,
for the much larger $a_c$ values used next, an evaluation of 
$U/(z J)$ was not possible since a VMC ground-state wave function
was not obtainable anymore for $a_c \ge 1$. This is because
VMC reweighting \cite{DuBois:01} breaks down at these large 
interactions. Instead, we resorted to evaluate the VPI SF fraction 
in each lattice well as was done in Sec.\ref{sec:superfluid-fraction}.

\begin{figure}[t!]
\includegraphics[width=8.5cm,bb = 75 544 526 775,clip]{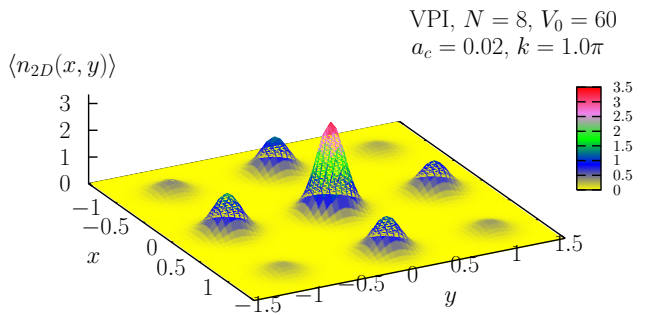}
\caption{Integrated VPI density along the $z-$axis 
$\langle n_{2D}(x,y)\rangle$ [Eq.(\ref{eq:n2D})] of a system with $N=8$, 
$a_c=0.02$, $k=1.0\pi$. The trapping geometry is the same as that of 
the systems in 
Fig.\ft\ref{fig:plot.n000_vs_ac.N8.B10.several.k.stack}, except 
that the OL potential barrier is $V_0 = 60$. The
$a_c$, $k$ and $V_0$ are in units of $a_{ho}$, $a_{ho}^{-1}$ and \hw, 
respectively, where \aho. The $x$ and $y$ are in units of $a_{ho}$, and
$\langle n_{2D}(x,y)\rangle$ is in units of $a_{ho}^{-2}$.} 
\label{fig:plot.od.dat.VPIac0.02N8B60k1.0}
\end{figure}

\begin{figure}[t!]
\includegraphics[width=8.5cm,bb=163 479 498 773,clip]{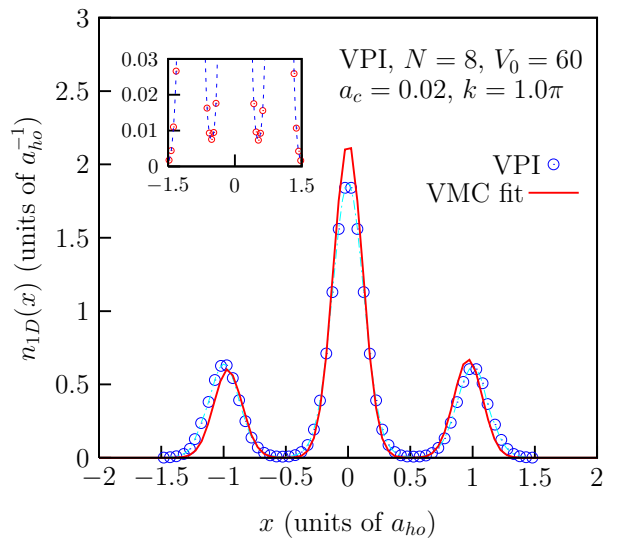}
\caption{Integrated VPI 1D optical density $\langle n_{1D}(x)\rangle$
[Eq.(\ref{eq:n1D})] 
of Fig.\ft\ref{fig:plot.od.dat.VPIac0.02N8B60k1.0} (open circles) 
along the $x-$axis. The solid red line is a VMC best ``fit" 
to the VPI data using the trial wave function 
Eq.(\ref{eq:trial-wave-function}). The inset is the same figure; but 
taken for a much smaller density range in order to reveal the 
overlap of the wave function between the lattice sites. The $a_c$, 
$k$ and $V_0$ are in units of $a_{ho}$, $a_{ho}^{-1}$ and \hw, 
respectively, where \aho.}
\label{fig:plot.density.slice.1D.ac0.02N8B60k1.0}
\end{figure}

\begin{figure*}[t!]
\includegraphics[width=17cm,bb = 66 490 565 775,clip]{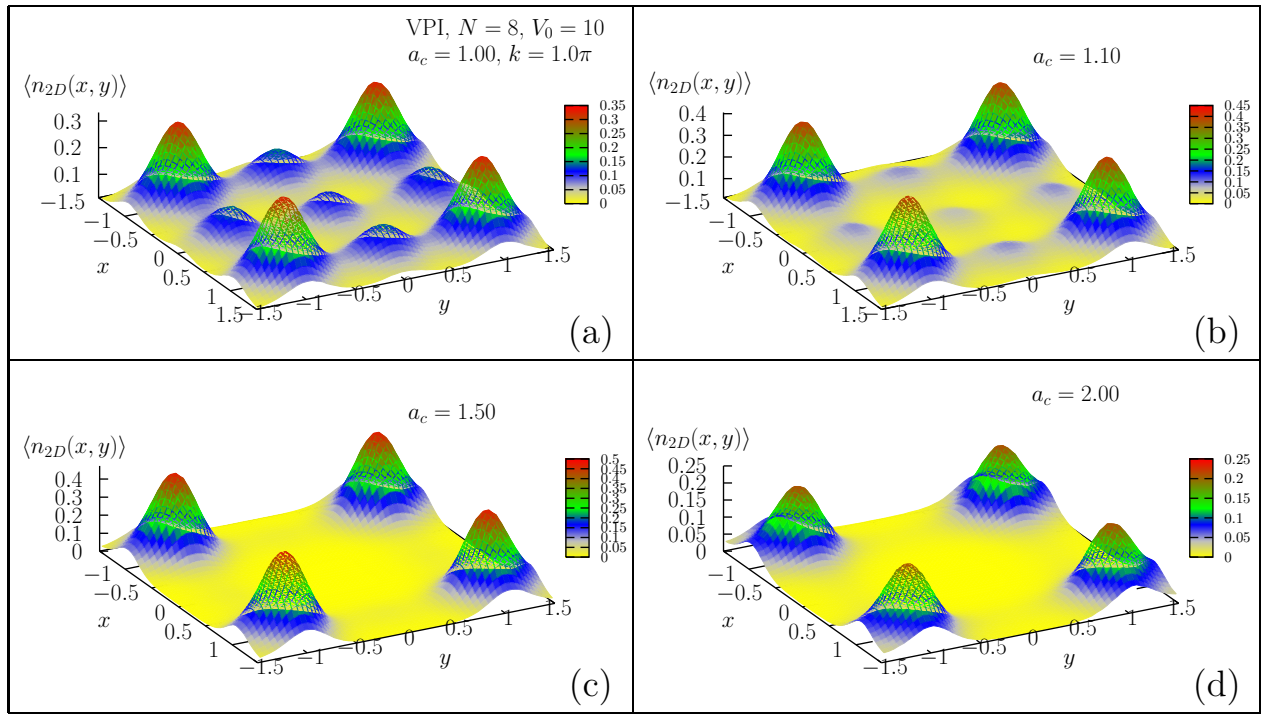}
\caption{As in Fig.\ft\ref{fig:plot.od.dat.VPIac0.02N8B60k1.0}; 
but for $V_0\,=\,10$ and (a) $a_c=1.00$, (b) 1.10, (c) 1.50,
and (d) 2.00, respectively. The $a_c$, $k$ and $V_0$ are in 
units of $a_{ho}$, $a_{ho}^{-1}$ and \hw, respectively, where \aho. 
The $x$ and $y$ are in units of $a_{ho}$, and $\langle n_{2D}(x,y)\rangle$ 
is in units of $a_{ho}^{-2}$.}
\label{fig:plot.od.dat.VPI.several.large.ac.N8.B10.k1.0.stack}
\end{figure*}

\subsubsection{Large $V_0$}

\hs First, we tried to obtain a {\it pure} MI by an increase of $V_0$. 
In Fig.\ft\ref{fig:plot.od.dat.VPIac0.02N8B60k1.0}, we show such
a case of high $V_0$. There, we display the integrated VPI 
density $\langle n_{2D}(x,y)\rangle$ [Eq.(\ref{eq:n2D})] for 
a system of $N=8$, $V_0 = 60$, $a_c = 0.02$, and $k = 1.0\pi$ 
in the same trapping geometry as that of the systems of 
Fig.\ft\ref{fig:plot.n000_vs_ac.N8.B10.several.k.stack}. The
corresponding Fig.\ft\ref{fig:plot.density.slice.1D.ac0.02N8B60k1.0} 
displays the integrated 1D VPI density [$\langle n_{1D}(x)\rangle$ 
(open circles), Eq.(\ref{eq:n1D})] of 
Fig.\ft\ref{fig:plot.od.dat.VPIac0.02N8B60k1.0} along the $x-$axis. 
The solid line is the integrated 1D VMC density obtained by
Eq.(\ref{eq:n1D}) as for VPI. It is a VMC best ``fit" to the VPI 
data, obtained by manually fitting the parameters of the 3D VMC trial 
function [Eq.(\ref{eq:trial-wave-function})] to the 1D VPI density. 
The goal was to use the ``fitted" VMC trial function to compute the ratio 
$\langle U_{(000)}\rangle/[6\langle J_{(000)\rightarrow(100)}\rangle]$ 
\cite{Sakhel:2010} between the central cell and a first neighbor cell 
in order to make a decision about the state of the system. It was found 
that the latter ratio equals 0.015 which is much less than 5.8,
placing the system of Fig.\ref{fig:plot.density.slice.1D.ac0.02N8B60k1.0} 
in the SF regime. We must also emphasize that the above ratio 
is difficult to obtain using VPI, as this method does not optimize
a parameterized trial function, but is rather independent of a 
trial function. Next to this, the VMC method failed to ``automatically"
optimize the trial wave function at this high value of $V_0$ in 
order to reach a ground-state. This is the reason why we resorted
to a ``manual" optimization. 

\hs The upper left inset of 
Fig.\ft\ref{fig:plot.density.slice.1D.ac0.02N8B60k1.0} shows a 
magnified view of the VPI density-overlap between the lattice wells.
In Fig.\ft\ref{fig:plot.od.dat.VPIac0.02N8B60k1.0}, one can
see that the tunneling of the BEC between the lattice wells is 
substantially reduced, however, upon a careful inspection of 
Fig.\ft\ref{fig:plot.density.slice.1D.ac0.02N8B60k1.0} and the 
inset, one can see that some overlap remains between the lattice 
wells. In fact, Eq.(\ref{eq:VPI-superfluid-fraction}) reveals a 
global SFF of $(23.93\,\pm\,0.06)\,\%$ for that system. That is, 
superfluidity is still present at this high $V_0$.

\subsubsection{Large $a_c$}

\begin{figure}[t!]
\includegraphics[width=8.5cm,bb=163 479 498 773,clip]{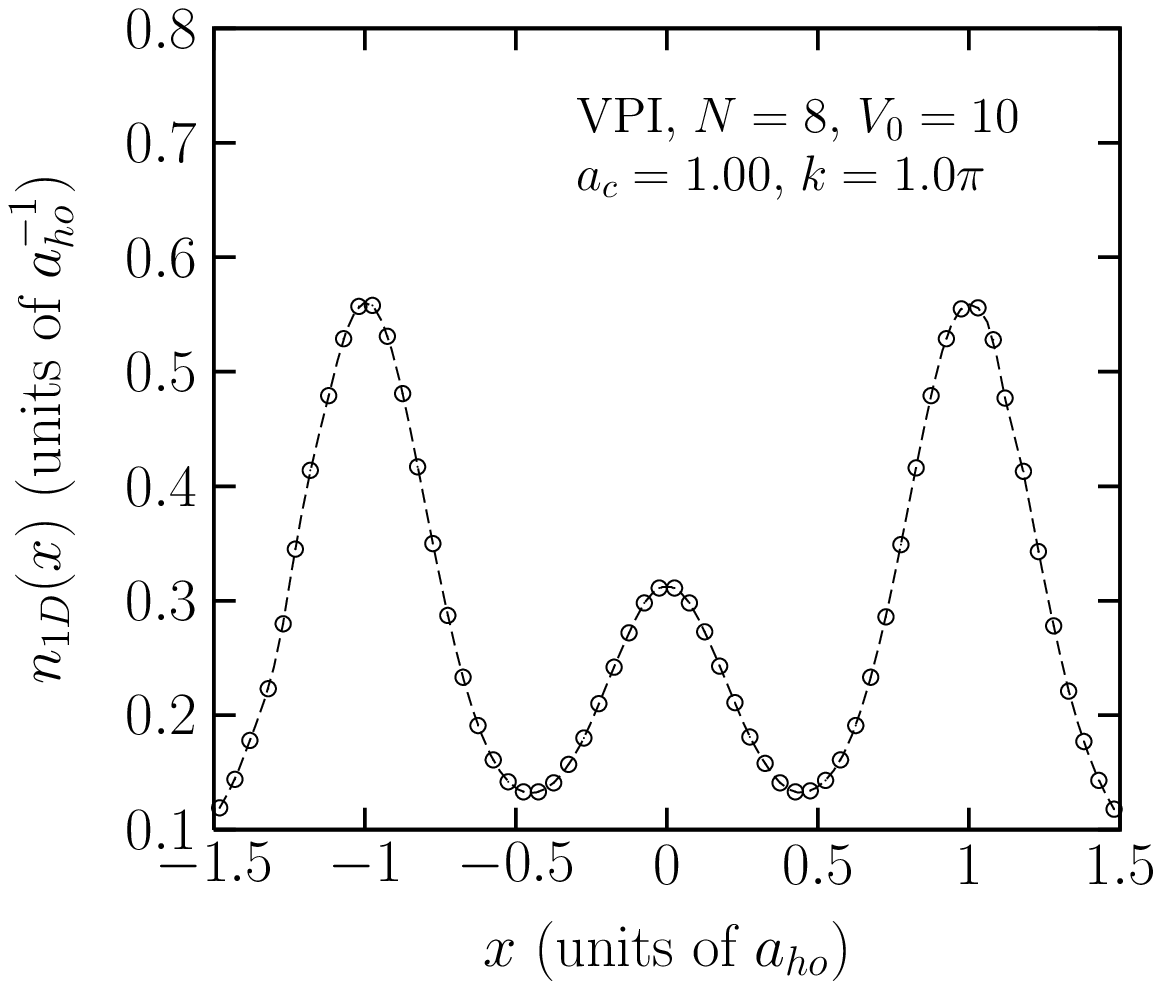}
\caption{Integrated VPI 1D optical density $\langle n_{1D}(x)\rangle$ 
[Eq.(\ref{eq:n1D})] of 
Fig.\ft\ref{fig:plot.od.dat.VPI.several.large.ac.N8.B10.k1.0.stack}(a) 
along the $x-$axis. The $a_c$, $k$ and $V_0$ are in units of $a_{ho}$, 
$a_{ho}^{-1}$ and \hw, respectively, where \aho.}
\label{fig:plot.density.slice.1D.ac1.00N8B10k1.0}
\end{figure}

\begin{figure}[t!]
\includegraphics[width=8.5cm,bb=163 479 498 773,clip]{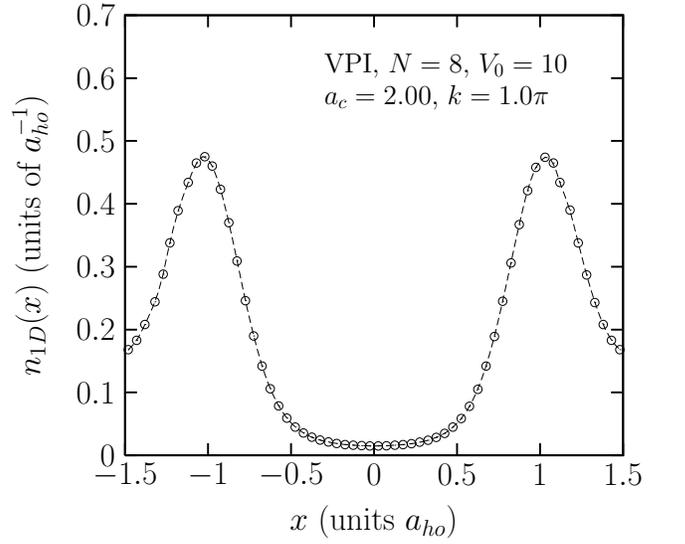}
\caption{Integrated VPI 1D optical density $\langle n_{1D}(x)\rangle$ 
[Eq.(\ref{eq:n1D})] of 
Fig.\ft\ref{fig:plot.od.dat.VPI.several.large.ac.N8.B10.k1.0.stack}(d) 
along the $x-$axis. The $a_c$, $k$ and $V_0$ are in units of $a_{ho}$, 
$a_{ho}^{-1}$ and \hw, respectively, where \aho.} 
\label{fig:plot.density.slice.1D.ac2.00N8B10k1.0}
\end{figure}

\hs Second, we tried to obtain a pure MI by increasing $a_c$ 
substantially. Therefore, 
Fig.\ft\ref{fig:plot.od.dat.VPI.several.large.ac.N8.B10.k1.0.stack},
displays several densities $\langle n_{2D}(x,y)\rangle$ for the
same system of Fig.\ft\ref{fig:plot.n000_vs_ac.N8.B10.several.k.stack}
with $k\,=\,1.0\pi$, obtained by increasing $a_c$ to extreme values: 
$a_c=1.00$ [frame (a)]; 1.10 [frame (b)]; 1.50 [frame (c)]; and 2.00 
[frame (d)], respectively. The density-peaks at the corners of the 
CHOCL have the largest amplitudes as the particles are repelled away 
from the center of the trap. At this $a_c$, the particles prefer to 
occupy the corners to reduce the repulsive potential energy, but
in (a) and (b) there is still some small probability for them to
occupy the wells closer to the center of the trap. It turns 
out that for (a) and (b) there remains BEC overlap (tunneling) between 
the lattice wells and their global SFF is $(46.31\,\pm\,0.07)\,\%$ 
for $a_c=1.00$ and ($39.82 \pm 0.31\%$) for $a_c=1.10$. The 
$\langle n_{1D}(x)\rangle$ of 
Fig.\ft\ref{fig:plot.od.dat.VPI.several.large.ac.N8.B10.k1.0.stack}(a)
displayed in Fig.\ft\ref{fig:plot.density.slice.1D.ac1.00N8B10k1.0} reveals
a remaining BEC overlap between the wells signalling the presence of 
superfluidity. For even larger $a_c$, notice that the central wells 
are almost vacant in both frames (c) and (d) of 
Fig.\ft\ref{fig:plot.od.dat.VPI.several.large.ac.N8.B10.k1.0.stack},
as the extremely strong repulsion has expelled all the atoms from 
the central cells. However, upon inspecting the $\langle n_{1D}(x)\rangle$ 
in Fig.\ft\ref{fig:plot.density.slice.1D.ac2.00N8B10k1.0} corresponding to 
Fig.\ft\ref{fig:plot.od.dat.VPI.several.large.ac.N8.B10.k1.0.stack}(d), 
one can observe for $a_c=2.00$ a still-existing BEC overlap between 
the lattice wells via the central cell, with a global SFF of 
$(22.46\,\pm\,0.05)\,\%$. 

\hs To this end, the main question that remains, then, is whether 
our systems above are only an SF, or a mixture of coexisting SF 
and MI regimes. One can possibly answer this question by computing the 
occupancies $\langle N_{(ijk)}\rangle$ of the individual lattice 
wells \cite{Rigol:2011}, using the counting method in Ref.\cite{Sakhel:2010}. 
If $\langle N_{(ijk)}\rangle$ is an integer, then one can talk about a 
MI. In that sense, Table \ref{table:occupancies} presents 
$\langle N_{(ijk)}\rangle$ obtained by VPI for the densities in 
Figs.\ft\ref{fig:plot.od.dat.VPIac0.02N8B60k1.0} and
\ref{fig:plot.od.dat.VPI.several.large.ac.N8.B10.k1.0.stack}(a)$-$(d).
In general, the $\langle N_{(ijk)}\rangle$ are all fractions and not 
integers, and $\langle N_{(000)}\rangle$, 
$\langle N_{(010)}\rangle$, and $\langle N_{(011)}\rangle\rightarrow 0$ 
for $a_c=1.50$ and 2.00, respectively. It might then be possible 
to argue that the empty cells
in Figs.\ref{fig:plot.od.dat.VPI.several.large.ac.N8.B10.k1.0.stack}(c) 
and (d) could constitute a empty (vacuum) MI regime. Therefore the 
CHOCL systems always retain a SF component coexistent with MI 
regimes.

\hs For a decisive check of the latter possibility, the local SFF
{\it with respect to the total} N, i.e. 
Eq.(\ref{eq:VPI-superfluid-fraction}) plus the condition 
(\ref{eq:SFinsideaCell}), was computed for the 
same systems of 
Fig.\ft\ref{fig:plot.od.dat.VPI.several.large.ac.N8.B10.k1.0.stack} at 
$a_c=1.50$ and 2.00. The reason for taking the SFF
with respect to $N$ instead of $\langle N_{(ijk)}\rangle$, is 
because the wells (000), (010), and (011) have a very low occupancy 
$\langle N_{(ijk)}\rangle$ for $a_c>1.20$; they are close to being 
empty. Therefore, it does not make sense to compute the SFF with 
respect to $\langle N_{(ijk)}\rangle$ as we did for the lower $a_c$
in Sec.\ref{sec:LSFF}. 

\begin{center}
\begin{table*}[ht]
\caption{VPI occupancies $\langle N_{(ijk)}\rangle$ of the systems in 
Figs.\ft\ref{fig:plot.od.dat.VPIac0.02N8B60k1.0} and
Figs.\ft\ref{fig:plot.od.dat.VPI.several.large.ac.N8.B10.k1.0.stack}(a-d),
plus an additional system at $a_c=1.20$ (density not shown), at various lattice 
sites (cells) $\mathbf{R}_n\equiv (ijk)$ representative of the whole OL. 
From left to right: $a_c$ is the HC diameter, $V_0$ is the OL depth,
followed by the cells (000), (010), (011), and (111), respectively. Lengths and energies are
in trap units \aho\ and \hw, respectively.}
\begin{tabular}{*{6}{|@{\hspace{0.5cm}} c@{\hspace{0.5cm}}}|}\hline
$a_c$ & $V_0$ & (000) & (010) & (011) & (111) \\ \hline
($a_{ho}$) & ($\hbar\omega_{ho}$) & \multicolumn{4}{c|}{} \\ \hline\hline
0.02 & 60 & 1.382                      & 0.536                       & 0.222                       & 0.058 \\ 
     &    & $\pm 3.169\times 10^{-3}$  & $\pm 2.449\times 10^{-3}$   & $\pm 1.797\times 10^{-3}$   & $\pm 9.103\times 10^{-4}$ \\ \hline
1.00 & 10 & 0.134                      & 0.125                       & 0.227                       & 0.471 \\ 
     &    & $\pm 1.338\times 10^{-3}$  & $\pm 1.101\times 10^{-3}$   & $\pm 1.446\times 10^{-3}$   & $\pm 1.936\times 10^{-3}$ \\ \hline
1.10 & 10 & 0.090                      & 0.075                       & 0.167                       & 0.597 \\
     &    & $\pm 1.255\times 10^{-3}$  & $\pm 8.760\times 10^{-4}$   & $\pm 1.270 \times 10^{-3}$  & $\pm 1.885\times 10^{-3}$ \\ \hline
1.20 & 10 & 0.015                      & 0.016                       & 0.064                       & 0.787                     \\
     &    & $\pm 3.723\times 10^{-4}$  & $\pm 2.732\times 10^{-4}$   & $\pm 5.350 \times 10^{-4}$  & $\pm 9.054\times 10^{-4}$ \\ \hline
1.50 & 10 & $6.429\times 10^{-4}$      & $2.063\times 10^{-3}$       & 0.027                       & 0.837                     \\
     &    & $\pm 7.495\times 10^{-5}$  & $\pm 8.418\times 10^{-5}$   & $\pm 4.349\times 10^{-4}$   & $\pm 9.144\times 10^{-4}$ \\ \hline 
2.00 & 10 & $1.386\times 10^{-6}$      & $7.444\times 10^{-5}$       & $5.472\times 10^{-3}$       & 0.583                     \\
     &    & $\pm6.864\times 10^{-7}$   & $\pm2.485\times 10^{-5}$    & $\pm1.585\times 10^{-4}$     & $\pm1.471\times 10^{-3}$     \\ \hline
\end{tabular}
\label{table:occupancies}
\end{table*}
\end{center}

\begin{table*}[ht]
\caption{VPI Superfluid fractions $(\rho_s/\rho)$ in individual cells 
[i.e., with respect to the {\it total} $N$ using 
Eq.(\ref{eq:VPI-superfluid-fraction}) plus the condition 
(\ref{eq:SFinsideaCell})] for some of the same systems in Table \ref{table:occupancies}.
Lengths and energies are in trap units \aho\ and \hw, respectively.}
\begin{tabular}{*{6}{|@{\hspace{0.5cm}} c@{\hspace{0.5cm}}}|}\hline
$a_c$ & $V_0$ & (000) & (010) & (011) & (111) \\ \hline
($a_{ho}$) & ($\hbar\omega$) & \multicolumn{4}{c|}{} \\ \hline\hline
1.50 & 10 & $0.000$    & $6.677\times 10^{-8}$    & $1.728\times 10^{-5}$    & $1.067\times 10^{-2}$ \\
     &    & $\pm0.000$ & $\pm2.380\times 10^{-8}$ & $\pm1.215\times 10^{-6}$ & $\pm1.742\times 10^{-5}$ \\ 
2.00 &    & $0.000$    & $2.005\times 10^{-7}$    & $4.820\times 10^{-7}$    & $4.878\times 10^{-3}$ \\
     &    & $\pm0.000$ & $\pm1.610\times 10^{-7}$ & $\pm1.520\times 10^{-7}$ & $\pm2.078\times 10^{-5}$ \\ \hline\hline
\end{tabular}
\label{table:superfluid-fractions-mixed-SF-MI-regimes}
\end{table*}

\hs Table \ref{table:superfluid-fractions-mixed-SF-MI-regimes}
presents results for $a_c>1.2$ and the same lattice wells as in 
Table \ref{table:occupancies}. Viewed from this angle, the SFF
is negligibly small in all lattice wells. There is a nonsignificant
SFF of $\sim 1\%$ for the well (111) at $a_c=1.50$, indicating
that some SF has migrated to the corners of the CHOCL. The lattice 
wells with a negligible SFF define a local MI phase. In contrast,
the global SFF is significant. 

\hs In conclusion, the HC Bose gases in the CHOCL traps presented
here develop coexisting SF-MI phases at extremely large HS radii
$a_c$. It also turns out, that one must carefully distinguish between
global and local superfluidity when it comes to defining local MI
domains. The presence of a global SF does not imply the absence of
a local MI phase. The systems presented here display a significant 
global SFF up to $a_c=2.00$; but the same cannot be stated about 
the local SFF in the lattice wells.

\subsection{Momentum density}\label{sec:momentum-density}

\hs In this section, we conclude with a computation of the 
$k-$space momentum distributions of ground-state densities.
The goal is to check for the presence of momentum states 
higher than $k=0$. According to a discussion by Mullin 
\cite{Mullin:1997}, Chester \cite{Chester:1968,Chester:1969} proved 
that there can be no BEC in $k>0$ states unless there is condensation 
into the $k=0$ momentum state. 

\begin{figure}[t!]
\includegraphics[width=8.0cm,bb=180 274 426 773,clip]{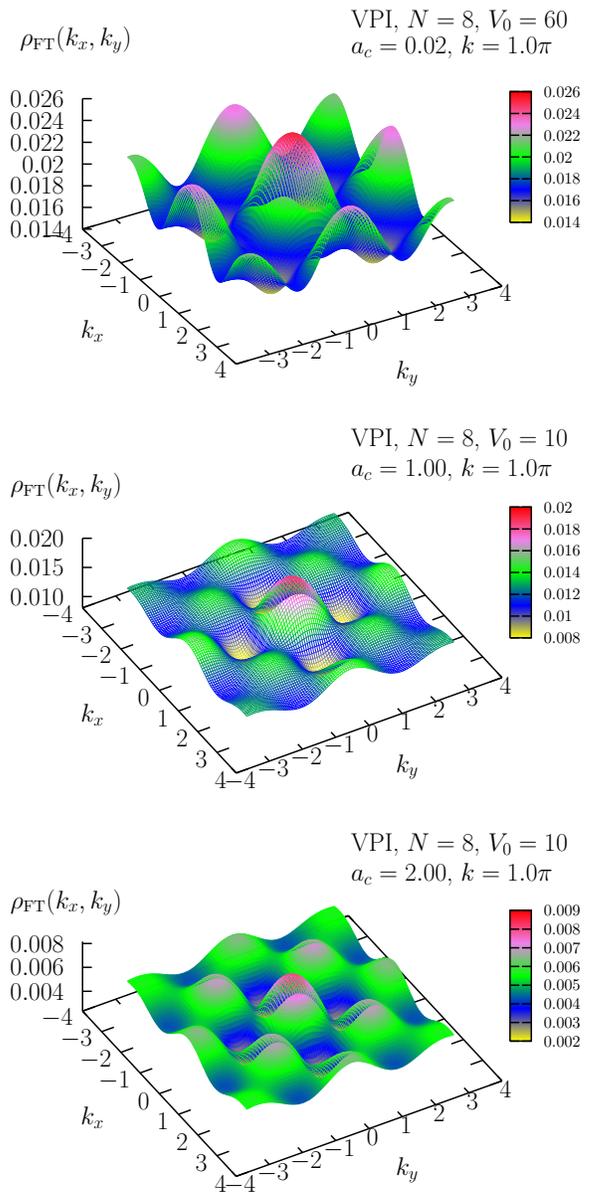}
\caption{Momentum densities obtained from Eq.(\ref{eq:momentum-density}).
Top frame: $\rho_{\hbox{\footnotesize FT}}(k_x,k_y)$ for 
Fig.\ft\ref{fig:plot.od.dat.VPIac0.02N8B60k1.0}; middle frame: for
Fig.\ft\ref{fig:plot.od.dat.VPI.several.large.ac.N8.B10.k1.0.stack}(a); 
bottom frame: for 
Fig.\ft\ref{fig:plot.od.dat.VPI.several.large.ac.N8.B10.k1.0.stack}(d).
The $\rho_{\hbox{\footnotesize FT}}(k_x,k_y)$ is in units of $a_{ho}^2$, 
and $k_x$ and $k_y$ are in units of $a_{ho}^{-1}$, where \aho. The
$a_c$, $k$ and $V_0$ are in units of $a_{ho}$, $a_{ho}^{-1}$ and \hw, 
respectively.}
\label{fig:plot.momd.N8.several.ac.stack}
\end{figure}

\hs The momentum density is calculated by a numerical Fourier 
transform (FT) of the spatial density according to

\begin{eqnarray}
&&\rho_{FT}(k_x,k_y)\,=\,\nonumber\\
&&\frac{1}{4\pi^2} \int_{-\infty}^{+\infty} dx 
\int_{-\infty}^{+\infty} dy \langle n_{2D}(x,y)\rangle 
\exp(-i \mathbf{k}\cdot\mathbf{r}), \nonumber\\
\label{eq:momentum-density}
\end{eqnarray}

where $\mathbf{r}\,=\,x \mathbf{i}\,+\,y\mathbf{j}$, and 
$\mathbf{k}\,=\,k_x\mathbf{i}\,+\,k_y\mathbf{j}$, where $\mathbf{i}$
and $\mathbf{j}$ are unit vectors.  
Fig.\ft\ref{fig:plot.momd.N8.several.ac.stack}
reveals $\rho_{FT}(k_x,k_y)$ from the top frame to the bottom frame,
respectively for Figs.\ft\ref{fig:plot.od.dat.VPIac0.02N8B60k1.0},
and \ref{fig:plot.od.dat.VPI.several.large.ac.N8.B10.k1.0.stack}(a) 
and (d). One observes that there is always a broad central BEC peak 
surrounded by Bragg peaks signalling the presence of $k>0$ states in 
the system. This is even the case for the system of 
Fig.\ref{fig:plot.od.dat.VPI.several.large.ac.N8.B10.k1.0.stack}(d), 
where no central spatial density peak is present. The $\rho_{FT}$ 
results are another manifestation of the Chester theorem mentioned 
above. 

\hs Next to this, had there been only one broad zero-momentum 
peak, this would have indicated the presence of a ``pure" MI state 
\cite{Hen:2009,Brouzos:2010}. The particles are therefore not locked 
in their positions as in a MI and rather display a mobility, arising 
from the hopping from one lattice well to another. Further, as a result 
of trapping, our central BEC density is not a sharp function of the 
momentum $\mathbf{k}$. This is in line with the finding of Hen and 
Rigol \cite{Hen:2010a}, who reported that the $\mathbf{k}=0$ density 
peak reveals a smoother dependence on $\mathbf{k}$ in a trapped system 
than a homogeneous system.

\hs Our momentum distributions in Fig.\ft\ref{fig:plot.momd.N8.several.ac.stack}
display similar features to Fig.1(a) of Spielman \ea\ \cite{Spielman:2007}.
The diffractive structure is indicative of the presence of a SF state 
(their Fig.17, top frame). As the system progresses into the MI regime, 
the diffractive structure weakens (their Fig.17, middle and bottom frames), 
as the intensity of the peaks declines. Accordingly, our results of 
Fig.\ft\ref{fig:plot.momd.N8.several.ac.stack} are in line with those of
Spielman \ea\ \cite{Spielman:2007} who indicated that (citing them:) ``the 
diffractive structure persists deep into the Mott regime". However, this
diffractive structure leads to other manifestations as outlined in the
next section. In addition, Gerbier \ea\ \cite{Gerbier:2005b} experimentally
explored phase coherence of ultracold Bose gases trapped in OLs.
These authors found that phase coherence persisted in the MI phase
by studying the interference pattern of the density distribution of an
expanding $^{87}$Rb BEC released from an OL. The persistence
of this interference in the MI phase was attributed to short-range
coherence fundamentally attributed to the presence of partice-hole pairs.

\hs According to Yi \ea\ \cite{Yi:2007}, the Bragg peaks in the 
momentum distribution should narrow with a rise of the repulsive 
interactions in the system. However, the momentum density in 
Fig.\ft\ref{fig:plot.momd.N8.several.ac.stack}
does not reveal this feature since even at extreme repulsion 
$a_c\sim O(1)$ the peaks have significant width. This is because 
the bosons are still able to hop between lattice sites to another 
generating a quasimomentum distribution \cite{Yi:2007}. The 
interactions have a finite range $a_c$ which significantly reduces 
the localization effect as $a_c$ becomes larger. It should be 
noted, that at very large $a_c$, such as $a_c=2.00$, the only 
way the bosonic HSs can move around is by ``turning around" each 
other. 

\hs Yi \ea\ also noted that an external harmonic trap broadens the 
momentum distribution of the bosons in an OL, particularly the 
condensate part. In fact, this is what is observed in 
Fig.\ft\ref{fig:plot.momd.N8.several.ac.stack}.

\subsection{Connecting with the work of Brouzos \ea\ \cite{Brouzos:2010}}

\hs In much relevance to our work, Brouzos \ea\ explored earlier
1D few bosonic systems in multiwell traps. They chiefly studied 
the effect of repulsive interactions on the distribution of 
particles in these traps. It was found, that for a 1D homogeneous 
multiwell trap with hard-wall boundaries and a commensurate 
filling factor (their) $\nu=1$, the wells become equally 
occupied as the repulsion between the bosons rises [their Fig.\ft 2(a)]. 
The particles thus tend to reduce their interaction energy by 
this rearrangement. In the weakly repulsive regime, and for a 
filling factor $\nu<1$, i.e. an incommensurate filling, the 
particles are driven away from the central wells towards the 
outer wells and do not equally occupy the wells 
[their Fig.\ft10(a)]. In analogy to the latter result, we find 
that for an (although) inhomogeneous 3D OL with filling $N/N_L<1$ 
(i.e. $\nu=N/N_L$), the particles are also driven away from the 
central well towards the outer lattice wells, particularly the 
corners of the CHOCL. The inset in their Fig.\ft10(b) shows how 
the populations of their wells change with the repulsion between 
the bosons. Most importantly, whereas the central site ($s=4$) 
gets less populated as their $g$ rises, the site ($s=1$) farther 
away gets more populated. Our 
Figs.\ft\ref{fig:plot.n000_vs_ac.N8.B10.several.k.stack} and 
\ref{fig:plot.nntilde_vs_ac.N8.B10.several.k.stack} display a 
similar behavior for the 3D CHOCL systems, although there does
not seem to be a transfer of particles in the intermediate lattice
wells between the central well of the CHOCL and its wells at the 
corners, as they reported. This is because none of the intermediate 
wells gains and subsequently loses population as those demonstrated by 
the inset of their Fig.\ft10(b).

\hs In contrast to their 1D inhomogeneous result with $\nu>1$ 
[their Fig.\ft14(a)], where even at large $g$ the central wells 
remains significantly populated, the central cell of the inhomogeneous 
3D CHOCL becomes almost vacant! Thus, there is a higher tendency 
for the central well to become vacant in a 3D CHOCL than a 1D 
multiwell trap with harmonic confinement. As such, the 
dimensionality of this system seems to play a crucial role in 
this specific feature. This is particularly since the particles 
in 3D do not face the ``energy obstacles" between the wells as 
much as in a 1D multiwell trap with harmonic confinement 
\cite{Brouzos:2010}, when the tunnel from the center to the 
corners of the CHOCL. These ``energetic obstacles" force the 
particles in a 1D inhomogeneous OL with $\nu>1$ to favor the 
occupation of the middle wells.

\hs For the filling factor $\nu<1$, their particles remained 
delocalized and coherence did not vanish completely, contrary 
to the commensurate filling case. Our CHOCL systems with $\nu<1$ 
are also incommensurately filled and as such will always display 
a delocalization of particles and an associated coherence manifested 
by a remaining SFF.

\hs With regards to their momentum distributions in Fig.\ft11(c), 
they display a structure with Bragg peaks, where the central peak 
is lowered because of a partial loss of coherence with increasing
repulsion between the bosons. Similarly, our 
Fig.\ft\ref{fig:plot.momd.N8.several.ac.stack} also displays a
lowering of the central peak with increasing repulsion. The 
Bragg peaks surrounding the central one in our
Fig.\ft\ref{fig:plot.momd.N8.several.ac.stack} reveal accordingly
incomplete localization and persisting coherence, even as the 
system enters into the MI regime. In the work of Brouzos \ea, 
the Bragg peaks of the momentum distribution of a uniform 1D Bose 
gas in a multiwell trap with $\nu=1$ vanish into a broad, smooth, 
Gaussian peak centered at $k=0$ with a rise of $g$ [their Fig.\ft5(a)].

\hs Most importantly, Brouzos \ea\ computed the CF in the lowest
natural orbital ($\ell=0$) [their Fig.\ft11(b)] and showed how it 
declines with increasing $g$, whereas it rises for higher ones
($\ell>0$). Our Figs.\ft\ref{fig:plot.n000_vs_ac.N8.B10.several.k.stack} 
and \ref{fig:plot.nntilde_vs_ac.N8.B10.several.k.stack} display the
same result for $\ell=0$, where for $(000)$, $(001)$, and $(010)$ 
the CF decreases. However, it rises for $(011)$ and $(111)$.

\hs Our results for the 3D CHOCL with $\nu\ne 1$ are then very 
much in line with those of Brouzos \ea\ \cite{Brouzos:2010} for
the 1D inhomogeneous case with $\nu\ne 1$.

\section{Conclusions}\label{sec:conclusions}

\hs In summary, then, we have presented a numerical investigation 
of the CF and SFF of a few HC bosons in a 3D CHOCL of $3\times 3\times 3$ 
lattice sites. The global and local CF and SFF were computed for 
each CHOCL well and were explored as functions of $a_c$ and the 
lattice wave vector $k=\pi/d$, $d$ being the lattice spacing. 
The role of the interference between the condensates in all 
lattice wells, and its effect on the CF in one well was also 
studied. In a major part of this work, an achievement of a ``pure" 
MI state was attempted, yet only mixed SF-MI phases were obtained.

\hs The most important result of this paper is an opposing behavior 
for the global CF and SFF as functions of $k$. Whereas the global 
CF increases with increasing $k$, the global SFF decreases with 
$k$. For the CF, this was explained by approximating each lattice
well by a HO trap and the preference of the bosons to occupy the
lowest HO level as the energy-level spacing increases with increasing
$k$. For the SF, this was explained by an increase in the localization
of the bosons in each lattice well as the local confinement strength
increases with increasing $k$.

\hs From a local perspective, it was found that the condensate 
is depleted with a rise of $a_c$ in the lattice wells 
$\mathbf{R}_n\equiv (000)$, (001), and (010), whereas it rises 
with increasing $a_c$ for (011) and (111). This is because of 
the tunneling of the condensate away from the trap center to the 
farther lattice sites at the edges of the trap. The local CF is 
also enhanced with a rise of $k$ in all lattice wells due to 
the same reasons outlined for the global CF. Further, when 
interference with all-neighbor sites is included, the CF in 
each lattice well is enhanced beyond the case with no interference 
effects. The local SFF as a function of $a_c$ in each lattice 
well displays a similar behavior to the CF as it overall declines 
in the wells (000) and (010) whereas it rises in (011) and (111). 
In contrast to the local CF, however, the local SFF decreases 
with increasing $k$, revealing again an opposing behavior to 
the CF on a local scale as well. An important point to mention,
is that the local CFs and SFFs are of the same order of magnitude. 

\hs The energy of the system rises with an increase in $a_c$, 
largely because the average onsite interactions rise at a rate 
overwhelming the drop in kinetic energy, or mobility of the bosons. 
Further, the energy does not tend to stabilize towards a 
fermionization limit as it happens in the 1D homogeneous multiwell
trap with commensurate filling $\nu=1$ \cite{Brouzos:2010}. 

\hs Finally, the possibility for achieving a ``pure" MI state in 
a 3D CHOCL with filling $\nu<1$ was investigated in order to compare
with the 1D homogeneous multiwell trap with hard-wall boundaries
explored by Brouzos \ea\ \cite{Brouzos:2010}. This was performed
by either increasing $V_0$ to 60 (\hw) or by increasing $a_c$ to
huge values $a_c \sim O(1)$. It was found that no ``pure" MI could
be achieved in a 3D CHOCL, but rather a coexisting SF-(vacuum)MI phase.
This is because for a filling $\nu < 1$, the particles remain 
delocalized and coherence cannot vanish completely even at huge
$a_c$. From another point of view, the small number of 
$3\times 3\times 3$ sites does not furnish the ground for a flat 
MI density \cite{Hen:2010b} to appear in the system. Consequently,
our systems remain superfluid, even in the extreme repulsive 
regime with a global SFF of $\sim 20\%$ for $a_c \ge 2$. The
presence of a global SFF does not guarantee the absence of local
MI regimes, however. Further, the empty lattice wells in the 
CHOCL define vacuum MI regimes.

\hs The momentum density of our systems with $\nu < 1$ reveals 
a broad central BEC peak surrounded by Bragg peaks, even at 
$a_c\sim O(1)$. This proves again the presence of a coexisting 
SF-MI phase in the extremely repulsive regime due to the 
delocalization of the particles. Thus, our results are in line 
with those presented by Brouzos \ea\ \cite{Brouzos:2010}. Had 
there been only one central BEC peak \cite{Hen:2009}, the 
systems would have been defined as a ``pure" MI.

\acknowledgments

\hs ARS is indebted to the University of Jordan for funding this research
project. Additional thanks go to William J. Mullin who earlier emphasized
the importance of investigating the CF and SFF of the systems in 
Ref.\cite{Sakhel:2010} and also for a critical reading
of the manuscript.

\bibliography{optical-lattice,MonteCarlo,wa,sfa}

\begin{thebibliography}{54}
\expandafter\ifx\csname natexlab\endcsname\relax\def\natexlab#1{#1}\fi
\expandafter\ifx\csname bibnamefont\endcsname\relax
  \def\bibnamefont#1{#1}\fi
\expandafter\ifx\csname bibfnamefont\endcsname\relax
  \def\bibfnamefont#1{#1}\fi
\expandafter\ifx\csname citenamefont\endcsname\relax
  \def\citenamefont#1{#1}\fi
\expandafter\ifx\csname url\endcsname\relax
  \def\url#1{\texttt{#1}}\fi
\expandafter\ifx\csname urlprefix\endcsname\relax\def\urlprefix{URL }\fi
\providecommand{\bibinfo}[2]{#2}
\providecommand{\eprint}[2][]{\url{#2}}

\bibitem[{\citenamefont{{R. Bach and K. Rzazewski}}(2004)}]{Bach:04}
\bibinfo{author}{\bibnamefont{{R. Bach and K. Rzazewski}}},
  \bibinfo{journal}{Phys. Rev. A} \textbf{\bibinfo{volume}{70}},
  \bibinfo{pages}{063622} (\bibinfo{year}{2004}).

\bibitem[{\citenamefont{{K. Xu, Y. Liu, D. E. Miller, J. K. Chin, W. Setiawan,
  and W. Ketterle}}(2006)}]{Xu:2006}
\bibinfo{author}{\bibnamefont{{K. Xu, Y. Liu, D. E. Miller, J. K. Chin, W.
  Setiawan, and W. Ketterle}}}, \bibinfo{journal}{Phys. Rev. Lett.}
  \textbf{\bibinfo{volume}{96}}, \bibinfo{pages}{180405}
  (\bibinfo{year}{2006}).

\bibitem[{\citenamefont{{K. Sun, C. Lannert, and S.
  Vishveshwara}}(2009)}]{Sun:2009}
\bibinfo{author}{\bibnamefont{{K. Sun, C. Lannert, and S. Vishveshwara}}},
  \bibinfo{journal}{Phys. Rev. A} \textbf{\bibinfo{volume}{79}},
  \bibinfo{pages}{043422} (\bibinfo{year}{2009}).

\bibitem[{\citenamefont{{Z. Chen and B. Wu}}(2010)}]{Chen:2010}
\bibinfo{author}{\bibnamefont{{Z. Chen and B. Wu}}}, \bibinfo{journal}{Phys.
  Rev. A} \textbf{\bibinfo{volume}{81}}, \bibinfo{pages}{043611}
  (\bibinfo{year}{2010}).

\bibitem[{\citenamefont{{A. Valizadeh, Kh. Jahanbani, and M. R.
  Kolahchi}}(2010)}]{Valizadeh:2010}
\bibinfo{author}{\bibnamefont{{A. Valizadeh, Kh. Jahanbani, and M. R.
  Kolahchi}}}, \bibinfo{journal}{Phys. Rev. A} \textbf{\bibinfo{volume}{81}},
  \bibinfo{pages}{023616} (\bibinfo{year}{2010}).

\bibitem[{\citenamefont{{R. Ramakumar and A. N. Das}}(2005)}]{Ramakumar:2005}
\bibinfo{author}{\bibnamefont{{R. Ramakumar and A. N. Das}}},
  \bibinfo{journal}{Phys. Rev. B} \textbf{\bibinfo{volume}{72}},
  \bibinfo{pages}{094301} (\bibinfo{year}{2005}).

\bibitem[{\citenamefont{{H. Pu, L. O. Baksmaty, W. Zhang, N. P. Bigelow, and P.
  Meystre}}(2003)}]{Pu:03}
\bibinfo{author}{\bibnamefont{{H. Pu, L. O. Baksmaty, W. Zhang, N. P. Bigelow,
  and P. Meystre}}}, \bibinfo{journal}{Phys. Rev. A}
  \textbf{\bibinfo{volume}{67}}, \bibinfo{pages}{043605}
  (\bibinfo{year}{2003}).

\bibitem[{\citenamefont{{P. J. Y. Louis, E. A. Ostrovskaya, C. M. Savage, and
  Y. S. Kivshar}}(2003)}]{Louis:2003}
\bibinfo{author}{\bibnamefont{{P. J. Y. Louis, E. A. Ostrovskaya, C. M. Savage,
  and Y. S. Kivshar}}}, \bibinfo{journal}{Phys. Rev. A}
  \textbf{\bibinfo{volume}{67}}, \bibinfo{pages}{013602}
  (\bibinfo{year}{2003}).

\bibitem[{\citenamefont{{N. Fabbri, D. Cl$\acute{e}$ment, L. Fallani, C. Fort,
  M. Modugno, K. M. R. van den Stam, and M. Inguscio}}(2009)}]{Fabbri:2009}
\bibinfo{author}{\bibnamefont{{N. Fabbri, D. Cl$\acute{e}$ment, L. Fallani, C.
  Fort, M. Modugno, K. M. R. van den Stam, and M. Inguscio}}},
  \bibinfo{journal}{Phys. Rev. A} \textbf{\bibinfo{volume}{79}},
  \bibinfo{pages}{043623} (\bibinfo{year}{2009}).

\bibitem[{\citenamefont{{S. Fang, R.-K. Lee, and D.-W.
  Wang}}(2010)}]{Fang:2010}
\bibinfo{author}{\bibnamefont{{S. Fang, R.-K. Lee, and D.-W. Wang}}},
  \bibinfo{journal}{Phys. Rev. A} \textbf{\bibinfo{volume}{82}},
  \bibinfo{pages}{031601(R)} (\bibinfo{year}{2010}).

\bibitem[{\citenamefont{{W. Yi., G.-D. Lin, and L.-M. Duan}}(2007)}]{Yi:2007}
\bibinfo{author}{\bibnamefont{{W. Yi., G.-D. Lin, and L.-M. Duan}}},
  \bibinfo{journal}{Phys. Rev. A} \textbf{\bibinfo{volume}{76}},
  \bibinfo{pages}{031602(R)} (\bibinfo{year}{2007}).

\bibitem[{\citenamefont{{D. van Oosten, P. van der Straten, and H. T. C.
  Stoof}}(2001)}]{vanOosten:2001}
\bibinfo{author}{\bibnamefont{{D. van Oosten, P. van der Straten, and H. T. C.
  Stoof}}}, \bibinfo{journal}{Phys. Rev. A} \textbf{\bibinfo{volume}{63}},
  \bibinfo{pages}{053601} (\bibinfo{year}{2001}).

\bibitem[{\citenamefont{{I. Hen, M. Iskin, and M. Rigol}}(2010)}]{Hen:2010b}
\bibinfo{author}{\bibnamefont{{I. Hen, M. Iskin, and M. Rigol}}},
  \bibinfo{journal}{Phys. Rev. B} \textbf{\bibinfo{volume}{81}},
  \bibinfo{pages}{064503} (\bibinfo{year}{2010}).

\bibitem[{\citenamefont{{M. Rigol, G. G. Batrouni, V. G. Rosseau, and R. T.
  Scalettar}}(2009)}]{Rigol:2009}
\bibinfo{author}{\bibnamefont{{M. Rigol, G. G. Batrouni, V. G. Rosseau, and R.
  T. Scalettar}}}, \bibinfo{journal}{Phys. Rev. A}
  \textbf{\bibinfo{volume}{79}}, \bibinfo{pages}{053605}
  (\bibinfo{year}{2009}).

\bibitem[{\citenamefont{{I. Hen and M. Rigol}}(2010)}]{Hen:2010a}
\bibinfo{author}{\bibnamefont{{I. Hen and M. Rigol}}}, \bibinfo{journal}{Phys.
  Rev. A} \textbf{\bibinfo{volume}{82}}, \bibinfo{pages}{043634}
  (\bibinfo{year}{2010}).

\bibitem[{\citenamefont{{J.-K. Xue, A.-X. Zhang, and J. Liu}}(2008)}]{Xue:2008}
\bibinfo{author}{\bibnamefont{{J.-K. Xue, A.-X. Zhang, and J. Liu}}},
  \bibinfo{journal}{Phys. Rev. A} \textbf{\bibinfo{volume}{77}},
  \bibinfo{pages}{013602} (\bibinfo{year}{2008}).

\bibitem[{\citenamefont{{R. B. Diener, Q. Zhou, H. Zhai, and T.-L.
  Ho}}(2007)}]{Diener:07}
\bibinfo{author}{\bibnamefont{{R. B. Diener, Q. Zhou, H. Zhai, and T.-L. Ho}}},
  \bibinfo{journal}{Phys. Rev. Lett.} \textbf{\bibinfo{volume}{98}},
  \bibinfo{pages}{180404} (\bibinfo{year}{2007}).

\bibitem[{\citenamefont{{I. Brouzos, S. Z\"ollner, and P.
  Schmelcher}}(2010)}]{Brouzos:2010}
\bibinfo{author}{\bibnamefont{{I. Brouzos, S. Z\"ollner, and P. Schmelcher}}},
  \bibinfo{journal}{Phys. Rev. A} \textbf{\bibinfo{volume}{81}},
  \bibinfo{pages}{053613} (\bibinfo{year}{2010}).

\bibitem[{\citenamefont{{I. B. Spielman, W. D. Philips, and J. V.
  Porto}}(2008)}]{Spielman:2008}
\bibinfo{author}{\bibnamefont{{I. B. Spielman, W. D. Philips, and J. V.
  Porto}}}, \bibinfo{journal}{Phys. Rev. Lett.} \textbf{\bibinfo{volume}{100}},
  \bibinfo{pages}{120402} (\bibinfo{year}{2008}).

\bibitem[{\citenamefont{{A. A. Shams and H. R. Glyde}}(2009)}]{Shams:09}
\bibinfo{author}{\bibnamefont{{A. A. Shams and H. R. Glyde}}},
  \bibinfo{journal}{Phys. Rev. B} \textbf{\bibinfo{volume}{79}},
  \bibinfo{pages}{214508} (\bibinfo{year}{2009}).

\bibitem[{\citenamefont{{D. Baillie and P. B. Blakie}}(2009)}]{Baillie:2009}
\bibinfo{author}{\bibnamefont{{D. Baillie and P. B. Blakie}}},
  \bibinfo{journal}{Phys. Rev. A} \textbf{\bibinfo{volume}{80}},
  \bibinfo{pages}{033620} (\bibinfo{year}{2009}).

\bibitem[{\citenamefont{{M. Greiner, O. Mandel, T. Esslinger, J. W. H{\"a}nsch,
  and I. Bloch}}(2002)}]{Greiner:02}
\bibinfo{author}{\bibnamefont{{M. Greiner, O. Mandel, T. Esslinger, J. W.
  H{\"a}nsch, and I. Bloch}}}, \bibinfo{journal}{Nature}
  \textbf{\bibinfo{volume}{415}}, \bibinfo{pages}{39} (\bibinfo{year}{2002}).

\bibitem[{\citenamefont{{O. Gygi, H. G. Katzgraber, M. Troyer, S. Wessel, and
  G. G. Batrouni}}(2006)}]{Gygi:06}
\bibinfo{author}{\bibnamefont{{O. Gygi, H. G. Katzgraber, M. Troyer, S. Wessel,
  and G. G. Batrouni}}}, \bibinfo{journal}{Phys. Rev. A}
  \textbf{\bibinfo{volume}{73}}, \bibinfo{pages}{063606}
  (\bibinfo{year}{2006}).

\bibitem[{\citenamefont{{R. Roth and K. Burnett}}(2003)}]{Roth:2003}
\bibinfo{author}{\bibnamefont{{R. Roth and K. Burnett}}},
  \bibinfo{journal}{Phys. Rev. A} \textbf{\bibinfo{volume}{67}},
  \bibinfo{pages}{031602(R)} (\bibinfo{year}{2003}).

\bibitem[{\citenamefont{{D. Tilahun, R. A. Duine, and A. H.
  MacDonald}}(2011)}]{Tilahun:2011}
\bibinfo{author}{\bibnamefont{{D. Tilahun, R. A. Duine, and A. H. MacDonald}}},
  \bibinfo{journal}{Phys. Rev. A} \textbf{\bibinfo{volume}{84}},
  \bibinfo{pages}{033622} (\bibinfo{year}{2011}).

\bibitem[{\citenamefont{{I. Hen and M. Rigol}}(2009)}]{Hen:2009}
\bibinfo{author}{\bibnamefont{{I. Hen and M. Rigol}}}, \bibinfo{journal}{Phys.
  Rev. B} \textbf{\bibinfo{volume}{80}}, \bibinfo{pages}{134508}
  (\bibinfo{year}{2009}).

\bibitem[{\citenamefont{{B. Capogrosso-Sansone, E. Kozik, N. Prokof'ev, and B.
  Svistunov}}(2007)}]{Sansone:07}
\bibinfo{author}{\bibnamefont{{B. Capogrosso-Sansone, E. Kozik, N. Prokof'ev,
  and B. Svistunov}}}, \bibinfo{journal}{Phys. Rev. A}
  \textbf{\bibinfo{volume}{75}}, \bibinfo{pages}{013619}
  (\bibinfo{year}{2007}).

\bibitem[{\citenamefont{{J. Li, Y. Yu, A. M. Dudarev, and Q.
  Niu}}(2006)}]{Li:06}
\bibinfo{author}{\bibnamefont{{J. Li, Y. Yu, A. M. Dudarev, and Q. Niu}}},
  \bibinfo{journal}{New J. Phys.} \textbf{\bibinfo{volume}{8}},
  \bibinfo{pages}{154} (\bibinfo{year}{2006}).

\bibitem[{\citenamefont{{M. J. Hartmann and M. B.
  Plenio}}(2008)}]{Hartmann:2008}
\bibinfo{author}{\bibnamefont{{M. J. Hartmann and M. B. Plenio}}},
  \bibinfo{journal}{Phys. Rev. Lett.} \textbf{\bibinfo{volume}{100}},
  \bibinfo{pages}{070602} (\bibinfo{year}{2008}).

\bibitem[{\citenamefont{{F. Gerbier, A. Widera, S. F\"olling, O. Mandel, T.
  Gericke, and I. Bloch}}(2005{\natexlab{a}})}]{Gerbier:05}
\bibinfo{author}{\bibnamefont{{F. Gerbier, A. Widera, S. F\"olling, O. Mandel,
  T. Gericke, and I. Bloch}}}, \bibinfo{journal}{Phys. Rev. A}
  \textbf{\bibinfo{volume}{72}}, \bibinfo{pages}{053606}
  (\bibinfo{year}{2005}{\natexlab{a}}).

\bibitem[{\citenamefont{{M. Yamashita and M. W. Jack}}(2007)}]{Yamashita:07}
\bibinfo{author}{\bibnamefont{{M. Yamashita and M. W. Jack}}},
  \bibinfo{journal}{Phys. Rev. A} \textbf{\bibinfo{volume}{76}},
  \bibinfo{pages}{023606} (\bibinfo{year}{2007}).

\bibitem[{\citenamefont{{M. Capello, F. Becca, M. Fabrizio, and S.
  Sorella}}(2007)}]{Capello:07}
\bibinfo{author}{\bibnamefont{{M. Capello, F. Becca, M. Fabrizio, and S.
  Sorella}}}, \bibinfo{journal}{Phys. Rev. Lett.}
  \textbf{\bibinfo{volume}{99}}, \bibinfo{pages}{056402}
  (\bibinfo{year}{2007}).

\bibitem[{\citenamefont{{S. Ramanan, T. Mishra, M. S. Luthra, R. V. Pai, and B.
  P. Das}}(2009)}]{Ramanan:2009}
\bibinfo{author}{\bibnamefont{{S. Ramanan, T. Mishra, M. S. Luthra, R. V. Pai,
  and B. P. Das}}}, \bibinfo{journal}{Phys. Rev. A}
  \textbf{\bibinfo{volume}{79}}, \bibinfo{pages}{013625}
  (\bibinfo{year}{2009}).

\bibitem[{\citenamefont{{A. Rancon and N. Dupuis}}(2012)}]{Rancon:2012}
\bibinfo{author}{\bibnamefont{{A. Rancon and N. Dupuis}}},
  \bibinfo{journal}{Phys. Rev. A} \textbf{\bibinfo{volume}{85}},
  \bibinfo{pages}{011602(R)} (\bibinfo{year}{2012}).

\bibitem[{\citenamefont{{G. G. Batrouni, V. Rosseau, R. T. Scalettar, M. Rigol,
  A. Muramatsu, P. J. H. Denteneer, and M. Troyer}}(2002)}]{Batrouni:2002}
\bibinfo{author}{\bibnamefont{{G. G. Batrouni, V. Rosseau, R. T. Scalettar, M.
  Rigol, A. Muramatsu, P. J. H. Denteneer, and M. Troyer}}},
  \bibinfo{journal}{Phys. Rev. Lett.} \textbf{\bibinfo{volume}{89}},
  \bibinfo{pages}{117203} (\bibinfo{year}{2002}).

\bibitem[{\citenamefont{{T. St\"oferle, H. Moritz, C. Schori, M. K\"ohl, and T.
  Esslinger}}(2004)}]{Stoeferle:2004}
\bibinfo{author}{\bibnamefont{{T. St\"oferle, H. Moritz, C. Schori, M. K\"ohl,
  and T. Esslinger}}}, \bibinfo{journal}{Phys. Rev. Lett.}
  \textbf{\bibinfo{volume}{92}}, \bibinfo{pages}{130403}
  (\bibinfo{year}{2004}).

\bibitem[{\citenamefont{{M. Snoek, I. Titvinidze, I. Bloch, and W.
  Hofstetter}}(2011)}]{Snoek:2011}
\bibinfo{author}{\bibnamefont{{M. Snoek, I. Titvinidze, I. Bloch, and W.
  Hofstetter}}}, \bibinfo{journal}{Phys. Rev. Lett.}
  \textbf{\bibinfo{volume}{106}}, \bibinfo{pages}{155301}
  (\bibinfo{year}{2011}).

\bibitem[{\citenamefont{{G. E. Astrakharchik and K. V.
  Krutitsky}}(2011)}]{Astrakharchik:2011}
\bibinfo{author}{\bibnamefont{{G. E. Astrakharchik and K. V. Krutitsky}}},
  \bibinfo{journal}{Phys. Rev. A} \textbf{\bibinfo{volume}{84}},
  \bibinfo{pages}{031604} (\bibinfo{year}{2011}).

\bibitem[{\citenamefont{{A. R. Sakhel, J. L. Dubois, and R. R.
  Sakhel}}(2010)}]{Sakhel:2010}
\bibinfo{author}{\bibnamefont{{A. R. Sakhel, J. L. Dubois, and R. R. Sakhel}}},
  \bibinfo{journal}{Phys. Rev. A} \textbf{\bibinfo{volume}{81}},
  \bibinfo{pages}{043603} (\bibinfo{year}{2010}).

\bibitem[{\citenamefont{{J. L. DuBois and H. R. Glyde}}(2001)}]{DuBois:01}
\bibinfo{author}{\bibnamefont{{J. L. DuBois and H. R. Glyde}}},
  \bibinfo{journal}{Phys. Rev. A} \textbf{\bibinfo{volume}{63}},
  \bibinfo{pages}{023602} (\bibinfo{year}{2001}).

\bibitem[{\citenamefont{{E. L. Pollock and D. M.
  Ceperley}}(1987)}]{Pollock:1987}
\bibinfo{author}{\bibnamefont{{E. L. Pollock and D. M. Ceperley}}},
  \bibinfo{journal}{Phys. Rev. B} \textbf{\bibinfo{volume}{36}},
  \bibinfo{pages}{8343} (\bibinfo{year}{1987}).

\bibitem[{Dub()}]{Dubois:2007}
\bibinfo{note}{{For this purpose, we modified a VPI code previously written by
  Jonathan L. DuBois which he has given us earlier.}}

\bibitem[{\citenamefont{{J. E. Cuervo, P.-N. Roy, and M.
  Boninsegni}}(2005)}]{Cuervo:05}
\bibinfo{author}{\bibnamefont{{J. E. Cuervo, P.-N. Roy, and M. Boninsegni}}},
  \bibinfo{journal}{J. Chem. Phys.} \textbf{\bibinfo{volume}{122}},
  \bibinfo{pages}{114504} (\bibinfo{year}{2005}).

\bibitem[{\citenamefont{{M. H. Kalos and P. A. Whitlock}}(1986)}]{Kalos:86}
\bibinfo{author}{\bibnamefont{{M. H. Kalos and P. A. Whitlock}}},
  \emph{\bibinfo{title}{{M}onte {C}arlo {M}ethods, {V}olume {I}: {B}asics}}
  (\bibinfo{publisher}{{{J}ohn {W}iley and {S}ons}}, \bibinfo{year}{1986}).

\bibitem[{\citenamefont{Arfken and Weber}(1995)}]{Arfken:1995}
\bibinfo{author}{\bibfnamefont{G.~B.} \bibnamefont{Arfken}} \bibnamefont{and}
  \bibinfo{author}{\bibfnamefont{H.~J.} \bibnamefont{Weber}},
  \emph{\bibinfo{title}{{Mathematical Methods for Physicists}}}
  (\bibinfo{publisher}{{Academic Press, San Diego}}, \bibinfo{address}{{USA}},
  \bibinfo{year}{1995}), \bibinfo{edition}{4th} ed.

\bibitem[{\citenamefont{{I. B. Spielman, W. D. Philips, and J. V.
  Porto}}(2007)}]{Spielman:2007}
\bibinfo{author}{\bibnamefont{{I. B. Spielman, W. D. Philips, and J. V.
  Porto}}}, \bibinfo{journal}{Phys. Rev. Lett.} \textbf{\bibinfo{volume}{98}},
  \bibinfo{pages}{080404} (\bibinfo{year}{2007}).

\bibitem[{\citenamefont{{A. R. Sakhel, J. L. DuBois, and H. R.
  Glyde}}(2008)}]{Sakhel:2008}
\bibinfo{author}{\bibnamefont{{A. R. Sakhel, J. L. DuBois, and H. R. Glyde}}},
  \bibinfo{journal}{Phys. Rev. A} \textbf{\bibinfo{volume}{77}},
  \bibinfo{pages}{043627} (\bibinfo{year}{2008}).

\bibitem[{\citenamefont{{William J. Mullin}}()}]{Mullin:2011}
\bibinfo{author}{\bibnamefont{{William J. Mullin}}}, \bibinfo{note}{{University
  of Massachusetts, Amherst MA, USA. Private communications}}.

\bibitem[{\citenamefont{{D. Jaksch, C. Bruder, J. I. Cirac, C. W. Gardiner, and
  P. Zoller}}(1998)}]{Jaksch:1998}
\bibinfo{author}{\bibnamefont{{D. Jaksch, C. Bruder, J. I. Cirac, C. W.
  Gardiner, and P. Zoller}}}, \bibinfo{journal}{Phys. Rev. Lett.}
  \textbf{\bibinfo{volume}{81}}, \bibinfo{pages}{3108} (\bibinfo{year}{1998}).

\bibitem[{\citenamefont{{Marcos Rigol}}()}]{Rigol:2011}
\bibinfo{author}{\bibnamefont{{Marcos Rigol}}}, \bibinfo{note}{{Department of
  Physics, Georgetown University, Washington DC, USA. Private communications.}}

\bibitem[{\citenamefont{{W. J. Mullin}}(1997)}]{Mullin:1997}
\bibinfo{author}{\bibnamefont{{W. J. Mullin}}}, \bibinfo{journal}{J. Low. Temp.
  Phys.} \textbf{\bibinfo{volume}{106}}, \bibinfo{pages}{615}
  (\bibinfo{year}{1997}).

\bibitem[{\citenamefont{{G. V. Chester}}(1968)}]{Chester:1968}
\bibinfo{author}{\bibnamefont{{G. V. Chester}}}, in
  \emph{\bibinfo{booktitle}{{Lectures in Theoretical Physics}}}, edited by
  \bibinfo{editor}{\bibnamefont{{K. T. Mahanthappa}}}
  (\bibinfo{publisher}{Gordon and Breach, Science Publishers, Inc., New York},
  \bibinfo{year}{1968}), p. \bibinfo{pages}{253}.

\bibitem[{\citenamefont{{G. V. Chester, M. E. Fisher, and N. D.
  Mermin}}(1969)}]{Chester:1969}
\bibinfo{author}{\bibnamefont{{G. V. Chester, M. E. Fisher, and N. D.
  Mermin}}}, \bibinfo{journal}{Phys. Rev.} \textbf{\bibinfo{volume}{185}},
  \bibinfo{pages}{760} (\bibinfo{year}{1969}).

\bibitem[{\citenamefont{{F. Gerbier, A. Widera, S. F\"olling, O. Mandel, T.
  Gericke, and I. Bloch}}(2005{\natexlab{b}})}]{Gerbier:2005b}
\bibinfo{author}{\bibnamefont{{F. Gerbier, A. Widera, S. F\"olling, O. Mandel,
  T. Gericke, and I. Bloch}}}, \bibinfo{journal}{Phys. Rev. Lett.}
  \textbf{\bibinfo{volume}{95}}, \bibinfo{pages}{050404}
  (\bibinfo{year}{2005}{\natexlab{b}}).

\end{thebibliography}

\end{document}